\documentclass[prd,nofootinbib,preprint]{revtex4}
\usepackage{graphicx}
\usepackage{gensymb}
\usepackage[T1]{fontenc} 
\usepackage{slashed}
\usepackage{epsfig}
\usepackage{amsmath}
\usepackage{amsfonts}
\usepackage{amssymb}
\usepackage{url}
\usepackage{hyperref}
\usepackage{subfigure}
\usepackage{hhline}
\usepackage{color}
\usepackage{bm}
\usepackage{cancel}
\newcommand{\beqa}{\begin{eqnarray}}
\newcommand{\eeqa}{\end{eqnarray}}
\newcommand{\beq}{\begin{equation}}
\newcommand{\eeq}{\end{equation}}

\newcommand{\nn}{\nonumber}
\newcommand{\bmt}{\begin{pmatrix}}
\newcommand{\emt}{\end{pmatrix}}
\usepackage[toc,page]{appendix}
\usepackage{comment}
\usepackage{hyperref}
\usepackage{epstopdf}
\usepackage{multirow}
\newcommand{\be}{\begin{equation}}
\newcommand{\ee}{\end{equation}}
\newcommand{\bea}{\begin{eqnarray}}
\newcommand{\eea}{\end{eqnarray}}

\begin{document}
\title{Exploring the effect of Lorentz invariance violation  with the currently running long-baseline experiments} 

\author{Rudra Majhi$^{1}$}
\email{rudra.majhi95@gmail.com}

\author{Soumya C.$^{2}$}
\email{soumyac20@gmail.com}

\author{Rukmani Mohanta$^{1}$}
\email{rmsp@uohyd.ac.in}

\affiliation{$^1$\,School of Physics, University of Hyderabad,
              Hyderabad - 500046, India \\                   
 $^2$\,Institute of Physics, Sachivalaya Marg,   Sainik School Post, Bhubaneswar 751005, India. }

\begin{abstract}
Neutrinos are the fundamental particles,  blind to all kind of interactions except the weak and gravitational. Hence, they can propagate very long distances without any deviation. This characteristic property can thus provide an ideal platform  to investigate Planck suppressed physics through their  long distance propagation. In this work,  we intend to investigate CPT violation through Lorentz invariance violation (LIV) in the  long-baseline accelerator based neutrino experiments.  Considering the simplest four-dimensional Lorentz violating parameters, for the first time, we obtain the sensitivity limits on the LIV parameters from the currently running long-baseline experiments T2K and NO$\nu$A. In addition to this, we show  their effects  on mass hierarchy and CP violation sensitivities by considering NO$\nu$A  as a case  study.  We find that the sensitivity limits on LIV parameters obtained from T2K are much weaker than that of NO$\nu$A  and the  synergy of T2K and NO$\nu$A  can improve these sensitivities. All these limits are slightly weaker ($2 \sigma$ level) compared to the values extracted from Super-Kamiokande experiment with atmospheric neutrinos.  Moreover,
we observe that the mass hierarchy  and CPV sensitivities are either enhanced or deteriorated significantly in the presence of LIV as these sensitivities  crucially depend  on the new CP-violating phases. We also present the correlation  between   $\sin^2 \theta_{23}$ and  the LIV parameter $|a_{\alpha \beta}|$,  as well as  $\delta_{CP}$ and $|a_{\alpha \beta}|$.

\end{abstract}
\maketitle
 
 \section{Introduction}
Neutrinos are considered to be the most fascinating particles in nature, posses many unique and interesting features in contrast to the other Standard Model (SM) fermions.  The effort of many dedicated neutrino oscillation experiments \cite{Fukuda:2001nj,Ahmad:2001an, Ahmad:2002jz, Ahmad:2002ka, Fukuda:1999pp, Fukuda:2000np, Apollonio:1997xe,An:2012eh,Abe:2011fz,Ahn:2012nd,Araki:2004mb,Eguchi:2002dm, Abe:2008aa,Agafonova:2018auq,Abe:2011sj,Adamson:2016tbq} over the last two decades, provide us a splendid understanding about the main features of these tiny and elusive particles.  Indeed we now know that neutrinos are massive albeit extremely light, and change their flavour as they propagate. This intriguing characteristic, known as neutrino oscillation, bestows the first experimental evidence of physics beyond the SM.  Without the loss of generality,  SM is considered as a low-energy effective theory,  emanating  from a fundamental unified picture of gravity and quantum physics at the Planck scale. To understand the nature  of the Plank scale physics through experimental signatures is therefore of great importance, though extremely challenging to identify.  Lorentz symmetry violation constitutes one of such signals,  basically  associated with tiny deviation from relativity. In recent times,  the search for Lorentz violating and related CPT violating signals have been explored over a wide range of systems and at remarkable sensitivities \cite{Kostelecky:2003cr,Arias:2007zz,Zhang:2018otj,Lang:2018yog,Aharmim:2018apr,Mewes:2019dhj,Samajdar:2019ptt,Martinez-Huerta:2019rkx,Huang:2019etr,Satunin:2019gsl,Katori:2019xpc,Quinn:2019ppv}.  One of the phenomenological consequences of CPT invariance is that a particle and its anti-particle will have exactly the same mass and lifetime and if any difference observed either in their mass or lifetime, would be a clear hint  for  CPT violation. There exists stringent experimental bounds on Lorentz and CPT violating parameters from  kaon and the lepton sectors.  For the kaon system, the observed mass difference  provides the upper limit on CPT violation as $\big | m_{K^0}-m_{\overline{K^0}} \big |/m_K <   6 \times 10^{-18}$ 
\cite{Tanabashi:2018oca}, which is quite stringent. However, parametrizing in terms of $m_K^2$ rather than $m_K$, as kaon is a boson and the natural mass parameter appears in the Lagrangian is the squared mass,  the kaon constraint turns out to be $\big | m_{K^0}^2-m_{\overline{K^0}}^2 \big |< 0.25 ~{\rm eV^2}$, which  is comparable to the   bounds obtained  from neutrino sector, though  relatively weak.
Furthermore, neutrinos are  fundamental particles, unlike the kaons 
 hence, the neutrino system can be regarded as a better  probe to search  for CPT violation.  For example, the current neutrino  oscillation data provides the most stringent bounds: $\big |\Delta m_{21}^2 -\Delta
\overline{m}_{21}^2 \big | < 5.9 \times 10^{-5}~{\rm eV}^2$ and $\big |\Delta m_{31}^2 -\Delta
\overline{m}_{31}^2 \big | < 1.1 \times 10^{-3}~{\rm eV}^2$ \cite{Ohlsson:2014cha}.
Recently, MINOS experiment \cite{Adamson:2013whj} has also provided  the bound on the atmospheric mass splitting for the neutrino and antineutrino modes at $3 \sigma$ C.L. as $|\Delta m_{31}^2 - \Delta \bar m_{31}^2| < 0.8 \times 10^{-3}~{\rm eV}^2$.  If these differences are due to the interplay of some kind of CPT violating new physics effects, they would influence  the oscillation phenomena for neutrinos and antineutrinos as well as have other phenomenological consequences, such as neutrino-antineutrino oscillation, baryogenesis \cite{Carmona:2004xc} etc.

It is well known that the local relativistic quantum field theories are based on three main ingredients:   Lorentz invariance, locality and hermiticity. The CPT violation is intimately related to Lorentz violation, as possible CPT violation can arise from Lorentz violation, non-locality, non-commutative geometry etc. 
So if CPT violation exists in nature and is related to quantum gravity, which is supposedly non-local and expected to be highly suppressed,   long-baseline experiments have the capability to probe such effects.  
Here, we present a brief illustration about, how the violation of Lorentz symmetry can affect the neutrino propagation.   In general, Lorentz symmetry breaking and quantum gravity are interrelated, which requires the existence of a universal  length scale for all frames.  However, such universal scale is in conflict with  general relativity, as length contraction is one of the consequences of Lorentz transformation. Such contradiction can be avoided by the modification of  Lorentz transformations (or in other words modifying dispersion relations).  The effects of perturbative Lorentz and CPT violation on neutrino oscillations has been studied in \cite{Diaz:2009qk}. Moreover, it has been shown explicitly in Ref. \cite{Barenboim:2018ctx}, how the oscillation probability  gets affected by the modified dispersion relation, however, for the sake of completeness we will present a brief discussion about it.  
The modified energy-momentum relation  for the neutrinos can be expressed as
\bea
E_i^2=p_i^2+ \frac{1}{2} m_i^2 \left (1+e^{2 A_i E_i/m_i^2} \right ),
\eea
 where $m_i$, $E_i$ and $p_i$ are  the mass, energy and momentum  of the $i$th neutrino in the mass basis, and $A_i$ is the dimensionful and Lorentz symmetry breaking parameter. Assuming  that all the neutrinos have the same energy ($E$), the  probability of transition from a given flavour $\alpha $  to another flavour $\beta$ for  two neutrino case is
 given as
 \bea
 P(\nu_\alpha \to \nu_\beta)=1-\sin^2 2 \theta \sin^2 \left (\frac{\Delta p L}{2} \right )\;,
 \eea 
 where $\theta$ represents the mixing angle and
 \bea
 \Delta p \approx \frac{\Delta m^2}{2 E} +\frac{1}{2} (A_i -A_j),
 \eea
 with $\Delta m^2= m_i^2-m_j^2$.
 Hence, the neutrino oscillation experiments might provide the opportunity to test this kind of new physics.  The limits on Lorentz and CPT violating parameters from MINOS experiment are presented in \cite{Rebel:2013vc}. The  possible effect of Lorentz violation in  neutrino oscillation phenomena has been intensely  investigated in recent years \cite{Diaz:2011ia, Antonelli:2018fbv, Diaz:2009qk, Arguelles:2019ifw, Aharmim:2018apr, Aartsen:2017ibm, Dai:2017sst, Katori:2016eni, Wei:2016ygk, Wang:2016lne, Abe:2014wla, Chatterjee:2014oda, Katori:2012hc,Diaz:2016xpw, Barenboim:2018ctx, Diaz:2014yva, Barenboim:2017ewj, Ge:2019tdi,Higuera:2016vcm, Abe:2012gw, Katori:2013jca, Agarwalla:2019rgv}.

In this paper, we are interested to study  the  phenomenological  consequences  introduced  in the neutrino sector due to the presence of Lorentz invariance violation terms.   In particular, we  investigate the impact of such new contributions  on the neutrino oscillation probabilities  for NO$\nu$A experiment. Further, we obtain the  sensitivity limits on the LIV parameters from the currently running long-baseline experiments T2K and NO$\nu$A. We also investigate the implications of  LIV effects on the determination of mass ordering as well as the CP violation discovery potential of NO$\nu$A experiment.

The outline of the paper is as follows. In section II, we present a brief discussion on the theoretical framework for incorporating LIV effects and their implications on neutrino oscillation physics. The simulation details used in this analysis are  discussed in section III. The impact of  LIV parameters on the $\nu_\mu \to \nu_e$  oscillation probability is presented in Section IV. Section V contains the discussion on the  sensitivity limits on LIV parameters, which can be  extracted from T2K and NO$\nu$A experiments. The discussion on how the discovery potential for CP violation and the mass hierarchy sensitivity get affected due to the presence of LIV, the correlation between LIV parameters and $\delta_{CP}$ as well as $\theta_{23}$ are illustrated in Section VI.  Finally we present our summary in section VII.

\section{Theoretical Framework} 
The Lorentz invariance violation effect can be introduced as a small perturbation to the standard physics descriptions of neutrino oscillations. Thus, 
 the effective Lagrangian that describes Lorentz violating neutrinos and anti-neutrinos \cite{Kostelecky:2003cr,Kostelecky:2011gq} is given as 
 \bea
 {\cal L}= \frac{1}{2} \bar \Psi_A(i \gamma^\mu{\partial}_\mu\delta_{AB}-M_{AB}+\hat{\cal Q}_{AB}) \Psi_B+
 {\rm h.c.}\;,
 \eea
  where  $\Psi_{A(B)}$ is a  $2N$ dimensional spinor containing the spinor field $\psi_{\alpha(\beta)}$ with $\alpha(\beta)$ ranges over $N$ spinor flavours and their charge conjugates $\psi_{\alpha(\beta)}^C=C\bar \psi_{\alpha(\beta)}^T$, expressed as $\Psi_{A(B)}= (\psi_{\alpha(\beta)}, \psi_{\alpha(\beta)}^C)^T$ and the Lorentz violating operator is characterized by  $\hat{\cal Q}$. Restricting ourselves to only  a renormalizable theory (incorporating terms with mass dimension $\leq$4), one can symbolically write  the Lagrangian   density for neutrinos as  \cite{Kostelecky:2011gq}  
 \bea
 {\cal L}_{\rm LIV}= -\frac{1}{2} \big [p_{\alpha \beta}^\mu \bar \psi_\alpha \gamma_\mu \psi_\beta +q_{\alpha \beta}^\mu \bar \psi_\alpha \gamma_5 \gamma_\mu \psi_\beta -i r_{\alpha \beta}^{\mu \nu} \bar \psi_\alpha  \gamma_\mu \partial_\nu \psi_\beta  -i s_{\alpha \beta}^{\mu \nu} \bar \psi_\alpha  \gamma_5 \gamma_\mu \partial_\nu \psi_\beta\big]+{\rm h.c.}\;,
 \eea
 where $p^\mu_{\alpha \beta}$, $q^\mu_{\alpha \beta}$, $r^{\mu \nu}_{\alpha \beta}$ and $s^{\mu\nu}_{\alpha \beta}$ are the Lorentz violating parameters,
 in the flavor basis.
 Since, only left-handed neutrinos are present in the SM, the observable effects which can be explored in the neutrino oscillation experiments can be parametrized as
 \bea
 (a_L)^\mu_{\alpha \beta}=(p+q)_{\alpha \beta}^\mu\;, ~~~~~~
  (c_L)^{\mu \nu}_{\alpha \beta}=(r+s)_{\alpha \beta}^{\mu \nu}\;. 
 \eea 
These parameters are hermitian matrices in the flavour space and can affect the standard vacuum Hamiltonian. The parameter $ (a_L)^\mu_{\alpha \beta}$ is related to  CPT 
 violating neutrinos and  $(c_L)^{\mu \nu}_{\alpha \beta}$ is associated with CPT-even, Lorentz violating neutrinos. Here, we consider the isotropic model (direction-independent) for simplicity, which appears when only the time-components of the coefficients are non-zero  i.e., terms with  $\mu=\nu=0$ \cite{Kostelecky:2003cr}. The sun-centred isotropic model is a popular choice and in this frame, the Lorentz-violating isotropic terms are considered as $(a)^0_{\alpha \beta}$ and $(c)^{00}_{\alpha \beta}$. 
  Here onwards we change the notation $(a_L)^0_{\alpha\beta}$ to $a_{\alpha\beta}$ and $(c_L)^{00}_{\alpha\beta}$ to $c_{\alpha\beta}$ for  convenience. 
 Taking into account only these isotropic terms of Lorentz violation parameters,  the  Hamiltonian for neutrinos, including LIV contributions becomes
 \bea
 H=H_{\rm vac}+H_{\rm mat}+H_{\rm LIV}\;,\label{Ham-LIV}
 \eea 
 where $H_{\rm vac}$  and $H_{\rm mat}$ correspond to the Hamiltonians in vacuum and in the presence of matter effects and $H_{\rm LIV}$ refers to the LIV Hamiltonian. These are  expressed as
 \bea 
 H_{\rm vac}&=& \frac{1}{2E} U
 \begin{pmatrix} m_1^2 & 0 & 0 \\
 0 & m_2^2 & 0\\
 0 & 0 & m_3^2\\
 \end{pmatrix}U^\dagger,~~~~~H_{\rm mat}= \sqrt 2 G_F N_e
 \begin{pmatrix} 1 & 0 & 0 \\
 0 & 0 & 0\\
 0 & 0 & 0\\
 \end{pmatrix},\\
 H_{\rm LIV}&=& \begin{pmatrix}
 a_{ee} & a_{e \mu} & a_{e \tau}\\
 a_{e\mu}^* & a_{\mu  \mu} & a_{\mu \tau}\\
  a_{e\tau}^* & a_{\mu  \tau}^* & a_{\tau \tau}\\
 \end{pmatrix}
 -\frac{4}{3}E
 \begin{pmatrix}
 c_{ee} & c_{e \mu} & c_{e \tau}\\
 c_{e\mu}^* & c_{\mu  \mu} & c_{\mu \tau}\\
  c_{e\tau}^* & c_{\mu  \tau}^* & c_{\tau \tau}\\
 \end{pmatrix}, \label{LIV-H}
 \eea
 where $U$ is the neutrino mixing matrix, $G_F$ is the Fermi constant and $N_e$ is the number density of electrons. The factor $-4/3$ in  $H_{\rm LIV}$  arises from the non-observability of the Minkowski trace of the CPT-even LIV parameter $c_L$, which forces the  $xx$, $yy$,  and $zz$ components to  be  related  to  the  00  component  \cite{Kostelecky:2003cr}. 
Since the mass dimensions of $a_{\alpha \beta}$ and $c_{\alpha \beta}$ LIV parameters are different, the effect of $a_{\alpha \beta}$ is proportional to the baseline $L$, whereas $c_{\alpha \beta}$ is proportional to $LE$ and  in this work 
we focus only on the impact of $a_{\alpha \beta}$ parameters on the physics potential of currently running long-baseline experiments NO$\nu$A and T2K.
Another possible way to introduce an isotropic Lorentz invariance violation is by considering the modified dispersion relation (MDR) preserving rotational symmetry \cite{Torri:2019gud}, which can be expressed as
  \bea
  E^2-\left (1-f\left(\frac{|\vec p|}{E} \right ) \right )|\vec p|^2=m^2,
  \eea
  where the perturbative function $f$ preserves the rotational invariance. However, this approach is not adopted in this work. 
  
It should be noted that, the Hamiltonian in the presence of LIV (\ref{Ham-LIV}), is analogous  to that in the presence of NSI in propagation, which is  expressed as 
\cite{Ohlsson:2012kf}
\bea
H=H_{\rm vac}+H_{\rm mat}+H_{\rm NSI}\;,
\eea
 with
 \bea
 H_{\rm NSI}= \sqrt 2 G_F N_e \begin{pmatrix}
 \epsilon_{ee}^m &  \epsilon_{e\mu}^m & \epsilon_{e\tau}^m\\
  \epsilon_{\mu e}^m &  \epsilon_{\mu \mu}^m & \epsilon_{\mu\tau}^m\\
    \epsilon_{\tau e}^m &  \epsilon_{\tau \mu}^m & \epsilon_{\tau\tau}^m\\
 \end{pmatrix}\;,
 \eea
 where   $\epsilon_{\alpha \beta}^m$ characterizes the relative strength between the matter effect due to NSI and the standard scenario. Thus, one obtains a correlation between the NSI and CPT violating scenarios through
 \bea
a_{\alpha \beta}= \sqrt 2 G_F N_e \epsilon_{\alpha \beta}^m \equiv V_{CC}\epsilon_{\alpha \beta}^m\;,
 \eea
 where $V_{CC}=\sqrt 2 G_F N_e$. The off-diagonal elements of the CPT violating LIV Hamiltonian ($a_{e \mu}$, $a_{e \tau}$ and $a_{\mu \tau}$) are the lepton flavor violating LIV parameters,  which can affect the neutrino flavour transition, are our subject of interest. These  parameters  are expected to be highly suppressed and the current limits on their values (in GeV), which are constrained by Super-Kamikande  atmoshperic neutrinos data at 95\% C.L.  \cite{Abe:2014wla} as
 \bea 
 |a_{e \mu}| <2.5 \times 10^{-23}\; ,~~~|a_{e \tau}|<5 \times 10^{-23}\; ,  ~~~~|a_{\mu \tau}|< 8.3 \times 10^{-24}\;.\label{sk}
 \eea 

\section{Simulation Details}
In this section, we briefly describe the experimental features of T2K and NO$\nu$A experiments that we consider in the analysis.\\ 
NO$\nu$A   is a currently running long-baseline accelerator experiment, with two totally active scintillator detectors, Near Detector (ND) and Far Detector (FD). ND is placed at around 1 km and FD is at a distance of 810 km away from source and both the detectors are off-axial by 14.6 mrad in nature, which provides a  large flux of neutrinos at an energy of 2 GeV,  the energy at which oscillation from $\nu_\mu$ to $\nu_e$  is expected to be at a maximum. It uses very high intensity $\nu_\mu$ beam, coming  from NuMI  beam of Fermilab, with beam power 0.7 MW and 120 GeV proton energy corresponding to  $6\times 10^{20}$ POT per year. This $\nu_\mu$ beam is detected by the  ND of mass 280 ton at Fermilab site and the oscillated neutrino beam is observed by 14 kton far detector located near Ash River. We assume 45\% (100\%) signal efficiencies for both electron (muon) neutrino and anti-neutrino signals. The background efficiencies for mis-identified muons (anti-muons) at the detector  as 0.83\% (0.22\%). The neutral current background efficiency for muon neutrino (antineutrino) is 2\% (3\%).  The background contribution coming from the existence of electron neutrino (anti-neutrino) in the beam, so called intrinsic beam contamination is about 26\% (18\%). Apart from these, we assume that 5\% uncertainty on signal normalization and 10\% on background normalization. The auxiliary files and experimental specification of NO$\nu$A experiment that we use for the analysis is taken from \cite{C.:2014ika}.

T2K (Tokai to Kamioka) experiment is making use of muon neutrino/anti-neutrino beam produced at Tokai  which is directed towards the detector of fiducial mass 22.5 kt kept 295 km far away at Kamioka \cite{Abe:2017uxa}. The detector is kept  2.5$^{\circ}$ off-axial to  the neutrino beam axis so that neutrino flux peaks around 0.6 GeV. To simulate T2K experiment, we consider the proton beam power of 750 kW and with  proton energy of 30 GeV which corresponds to a total exposure of 7.8 $\times 10^{21}$ protons on target (POT) with 1:1 ratio of neutrino to anti-neutrino modes. We match the signal and back-ground event rates  as given in the latest publication of the T2K collaboration \cite{Abe:2014tzr}. We consider an uncorrelated 5\% normalization error on signal and 10\% normalization error on background for both the appearance and disappearance channels as given in reference \cite{Abe:2014tzr} for both the neutrino and anti-neutrino.  We use the  Preliminary Earth Reference Matter (PREM) profile to calculate line-averaged constant Earth matter density ($\rho_{\rm avg}$=2.8 g/cm$^3$) for both NO$\nu$A and T2K experiments.

We use GLoBES software package along with snu plugin \cite{Huber:2004gg,Huber:2009cw} to simulate the experiments. The implementation of LIV  in neutrino oscillation scenario has been done  by modifying the snu code in accordance with the Lorentz violating Hamiltonian (\ref{Ham-LIV}). We use the values of standard three flavor oscillation parameters as given in  Table \ref{P.table} and  consider one LIV parameter at a time, while setting all other  parameters to zero unless otherwise mentioned. As mentioned before, we have considered only the isotropic CPT violating parameters ($a_{\alpha\beta}$) for our analysis. 
The values of the LIV parameters considered in our analysis are: $|a_{e \mu}|=|a_{\mu \tau}|=|a_{e \tau}|=2 \times 10^{-23}$ GeV and  $|a_{e e}|=$
$|a_{\mu \mu}|=|a_{\tau \tau}|=1 \times 10^{-22}$ GeV.
\begin{table}
\begin{tabular}{|c|c|c|} \hline
Parameter            & True value              & Marginalization Range  \\ \hline
$\sin^2 \theta_{12}$  & 0.310        & Not marginalized \\ \hline
$\sin^2 \theta_{13}$ & 0.0224                    & Not marginalized \\  \hline
$\sin^2 \theta_{23} $ & 0.5                       & $[0.4, 0.6]$\\ \hline
$\delta_{CP} $       & $ -\pi/2$                  & $[-\pi,\pi]$\\ \hline
$\Delta m^2_{21}$    & $7.39 \times 10^{-5}{\rm eV}^2 $ & 
Not marginalized \\ \hline

$\Delta m^2_{31}$    & $ 2.5 \times 10 ^{-3}{\rm eV}^2$& $[2.36 , 2.64] \times 10^{-3}{\rm eV}^2$  \\  \hline

\end{tabular}
\caption{The values of oscillation parameters that we consider in our analysis \cite{Esteban:2018azc}.}
\label{P.table}
\end{table}
 \section{Effect of LIV parameters  on $\nu_{\mu} \to \nu_e$ and $\nu_{\mu} \to \nu_{\mu}$   Oscillation Channels}
 In this section, we discuss the effect of LIV parameters $a_{\alpha \beta}=|a_{\alpha \beta}| e^{i \phi_{\alpha \beta}},(\phi_{\alpha \beta}=0$, for $\alpha=\beta)$, on  $\nu_{\mu} \to \nu_e$ oscillation channel, as the long-baseline experiments are  mainly looking at this oscillation channel. The evolution equation for a neutrino state $|\nu\rangle  = (|\nu_e\rangle, |\nu_\mu \rangle, |\nu_\tau \rangle)^T$, travelling a distance $x$, can be expressed as
 \bea
 i \frac{d }{dx} |\nu\rangle = H |\nu\rangle,
 \eea
 where $H$ is the effective Hamiltonian given in Eq. (\ref{Ham-LIV}). Then the oscillation probability for the transition $\nu_\alpha \to \nu_\beta $, after travelling a distance $ L$ can be obtained as is 
 \bea
 P_{\alpha \beta}= \left | \langle \nu_\beta| \nu_{\alpha}(L)\rangle \right |^2=\left |  \langle \nu_\beta| e^{-i H L} |\nu_{\alpha}\rangle \right |^2. \label{Prob}
\eea 
 Neglecting higher order terms, the oscillation probability  for $\nu_\mu\to \nu_e$  channel  in the presence of  LIV for NH can be expressed, which is  analogous to the NSI case as
 \cite{Liao:2016hsa,Kopp:2007ne,Chatterjee:2018dyd, Chaves:2018sih,Deepthi:2016erc,Deepthi:2017gxg,Dey:2018yht,Yasuda:2007jp,Masud:2015xva,Masud:2016gcl,Masud:2016bvp}, 
 
\begin{eqnarray}
P_{\mu e}^{\rm LIV}&\simeq&x^2 f^2 +2 x y fg \cos (\Delta+\delta_{CP}) +y^2 g^2+4 r_A |a_{e \mu}|\big\{ xf \big [ f s_{23}^2 \cos (\phi_{e \mu}+\delta_{CP})\nn\\
&&~~~~+g c_{23}^2 \cos( \Delta +\delta_{CP}+\phi_{e \mu})\big]+yg \big[ gc_{23}^2 \cos \phi_{e \mu} +f s_{23}^2\cos (\Delta-\phi_{e \mu})\big ]\big \}\nn\\
&+& 4 r_A |a_{e\tau}| s_{23} c_{23} \big\{xf \big[ f \cos (\phi_{e \tau}+\delta_{CP})- g\cos (\Delta +\delta_{CP}+\phi_{e \tau}) \big] \nn\\
&-&yg[g \cos \phi_{e \tau} -f \cos(\Delta-\phi_{e\tau})\big]\big\}+ 4 r_A^2 g^2 c_{23}^2 | c_{23} |a_{e \mu}| -s_{23} |a_{e\tau}||^2\nn\\
&+& 4 r_A^2 f^2 s_{23}^2  | s_{23} |a_{e \mu}| +c_{23} |a_{e\tau}||^2 +
8 r_A^2 f g s_{23}c_{23} \big\{ c_{23} \cos \Delta \big[s_{23}(|a_{e \mu}|^2 -|a_{e \tau}|^2) \nn\\
&+& 2 c_{23} |a_{e \mu}| |a_{e \tau}| \cos(\phi_{e \mu}-\phi_{e \tau}) \big]-|a_{e \mu}|| a_{e \tau}| \cos (\Delta -\phi_{e \mu} +\phi_{e \tau})\big \} +{\cal O}(s_{13}^2 a, s_{13}a^2, a^3)\;,\nn\\\label{pmue}
\end{eqnarray} 
 where
 \bea
&& x=2 s_{13}s_{23}\;,~~y=2rs_{12}c_{12}c_{23}\;,~~r=|\Delta m^2_{21}/\Delta m^2_{31}|\;,~~\Delta = \frac{\Delta m^2_{31} L}{4E}\;,~~V_{CC}=\sqrt 2 G_F N_e\nn\\
&& f=\frac{\sin\big[\Delta (1-r_A(V_{CC}+a_{ee}))  \big]}{1-r_A(V_{CC}+a_{ee})}\;,~~~~g=\frac{\sin\big[\Delta r_A(V_{CC}+a_{ee})  \big]}{r_A(V_{CC}+a_{ee})}\;,~~~r_A=\frac{2E}{{\Delta m}^2_{31}}\;,\hspace{0.5 true cm}\label{os-po}
 \eea
and  $s_{ij}=\sin\theta_{ij}$, $c_{ij}=\cos\theta_{ij}$. The antineutrino probability $ P_{\bar \mu \bar e}^{\rm LIV}$ can be obtained from (\ref{pmue}) by replacing $V_{CC} \to -V_{CC}$, $\delta_{CP} \to - \delta_{CP}$ and 
 $a_{\alpha \beta} \to 
-a_{\alpha \beta}^*$. Similar expression for inverse hierarchy can be obtained by substituting $\Delta m_{31}^2 \to - \Delta m_{31}^2$, i,e., $\Delta \to -\Delta$ and $r_A \to - r_A$.
 One can notice  from  Eq. (\ref{pmue}), that only the LIV parameters $a_{ee}$, $a_{e\mu}$ and $a_{e\tau}$  contribute to  appearance probability expression at leading order and the rest of the  parameters appear only on  sub-leading terms.
Since Eq. (\ref{pmue}) is valid only for small non-diagonal  LIV parameter $a_{\alpha \beta}$, in our simulations the oscillation probabilities are evaluated using Eq. (\ref{Prob}) without any such approximation, by modifying the neutrino oscillation probability function inside snu.c and implementing the Lorentz violating Hamiltonian (\ref{Ham-LIV}).

The expression for the survival  probability for the transition $\nu_\mu \to \nu_\mu$, up to ${\cal {O}}(r,s_{13},a_{\alpha \beta})$   is \cite{Chatterjee:2018dyd},
\begin{eqnarray} \nn
&&P_{\mu\mu}^{\rm LIV}\simeq 1- \sin^2 2 \theta_{23}\sin ^2 \Delta  \\ \nonumber
&&~~~~~~~- |a_{\mu\tau}|\cos \phi_{{\mu\tau}} \sin 2 \theta_{23}\Big[ (2r_A\Delta )\sin^2 2\theta_{23}\sin 2\Delta + 4 \cos^2 2 \theta_{23}r_A\sin ^2\Delta\Big]\\ 
&&~~~~~~~+(|a_{\mu\mu}|-|a_{\tau\tau}|)\sin^2  2 \theta_{23} \cos 2 \theta_{23}\Big[(r_A\Delta) \sin 2\Delta  -2r_A \sin ^2 \Delta  \Big].
\label{pmumu}
\end{eqnarray}
 It is important to observe from the  survival probability expression (\ref{pmumu}) that, the LIV parameters involved in $\nu_{\mu}\to \nu_{e}$ transitions do not take part in $\nu_{\mu}\to \nu_{\mu}$ channel. This probability  depends only  on the new parameters $a_{\mu\mu},|a_{\mu\tau}|$, $\phi_{\mu\tau}$ and $a_{\tau\tau}$. 
\begin{figure}
\includegraphics[scale=0.45]{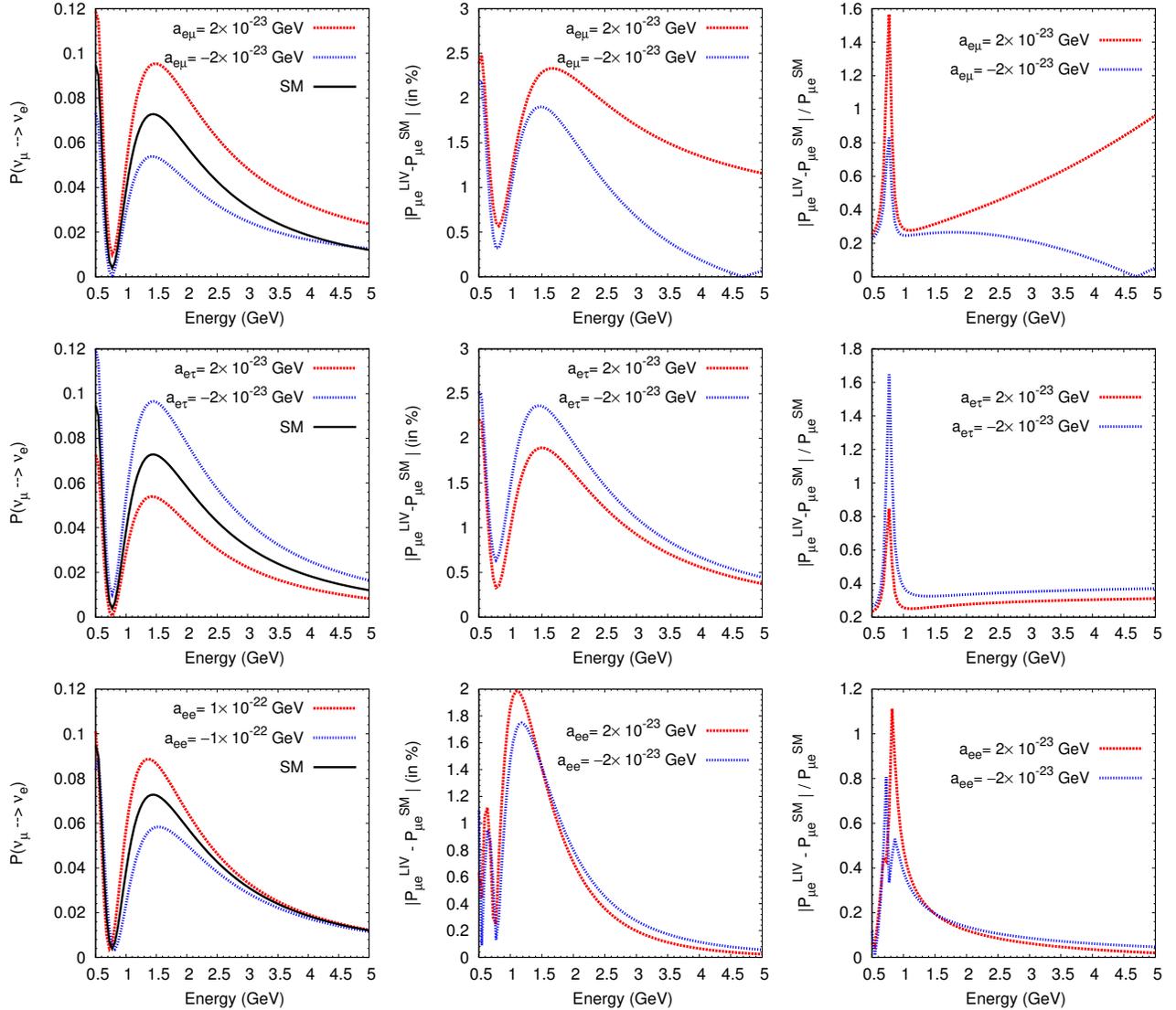}
\caption{ The   numerical oscillation probabilities for $\nu_e$ appearance channel  as a function of neutrino energy for NO$\nu$A experiment, in  presence of Lorentz violating parameters  $a_{e\mu}$, $a_{e\tau}$  and $a_{ee}$  in the left panel. The difference in the oscillation probabilities (in \%) with and without LIV are shown in the middle panel whereas the relative change in probabilities  are in the right panel.}
\label{pro-nd}
\end{figure}

\begin{figure}
\includegraphics[scale=0.45]{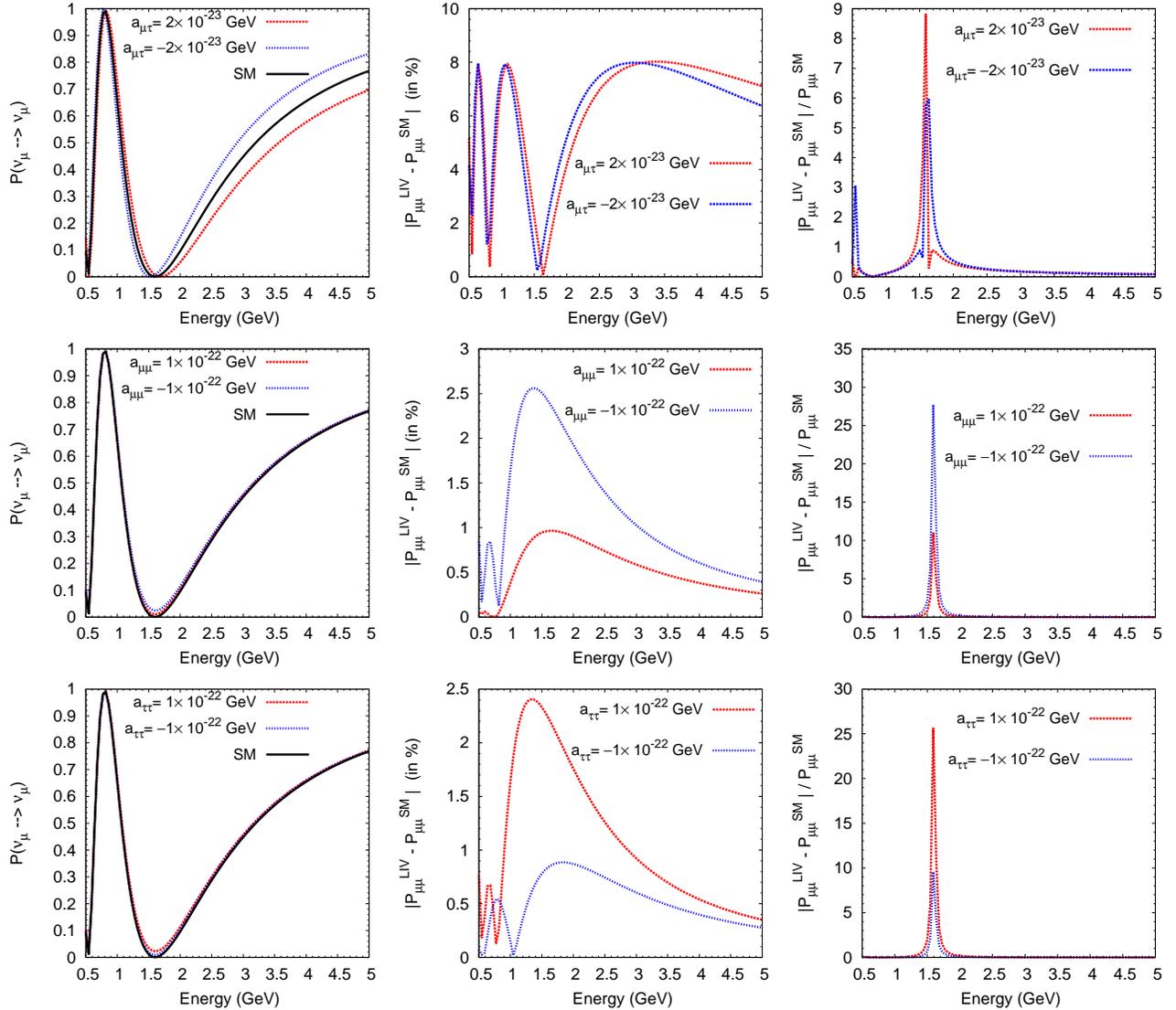}
\caption{Same as Fig.1 for the  $\nu_{\mu}$ survival probabilities as a function of  neutrino energy in presence of $a_{\mu\mu}$, $a_{\mu\tau}$, and $a_{\tau\tau}$ LIV parameters  for NO$\nu$A experiment.}
\label{pro-d}
\end{figure}

 The effect of 
LIV parameters on  $\nu_\mu \rightarrow \nu_e$ channel  for NO$\nu$A experiment is  displayed in  Fig. \ref{pro-nd}. 
The left panel of the figure shows how the oscillation probability gets modified in presence of LIV,  the absolute difference of standard case from Lorentz violating case (in \%) is shown in the middle panel and the relative change of the probability  $\frac{|P_{\alpha\beta}^{\rm LIV}-P_{\alpha\beta}^{SM}|}{P_{\alpha\beta}^{SM}}$ is shown in the right panel of the figure. In each plot, the black curve  corresponds to oscillation probability in the standard three flavor oscillation paradigm and red (blue) dotted curve  corresponds to the oscillation probability in presence of  LIV parameters with positive (negative) value.
 From  Fig. \ref{pro-nd}, it is clear that 
 all the three $a_{e\mu}$, $a_{e\tau}$ and $a_{ee}$ LIV parameters have significant impact on  the oscillation probability.
 It should be further noted that the parameters $a_{e\tau}$ and $a_{e\mu}$ have impact on the amplitude of oscillation and $a_{ee}$ is affecting to phase of the oscillation, which can be seen from the Eq. (\ref{pmue}). 
  It should   be  noted from the figure that positive and negative values for LIV parameter $a_{e\tau}$,   shift the probabilities in opposite direction of the standard probability curve,  while the case of $a_{e\mu}$ is just opposite to that of $a_{e\tau}$ and it also creates a distortion on the probability.  Also as seen from the  right panel of the Fig.\ref{pro-nd}, the relative change of the probability for LIV case with respect to the standard case, becomes significant towards lower energy. 
Furthermore, it should  be inferred from the left panel of the figure that the positive and negative values of LIV parameters affect the oscillation probabilities differently. However, the result is qualitatively independent of the actual sign of LIV parameters, i.e., the spectral form of the probability is same as the standard case both for positive and negative values of LIV parameters, either it is enhanced or reduced with respect to the standard oscillation probability.   Hence, one can take the $|a_{\alpha\beta}|$ for sensitivity study of the experiment in presence of LIV parameters. In Fig. \ref{pro-d}, the effect of LIV parameters $a_{\mu\mu}$, $a_{\mu\tau}$, and $a_{\tau\tau}$ on $\nu_{\mu}$ survival probability is  displayed. Analogous to the previous case, here also the effects of the  parameters are noticeable; the parameter $|a_{\mu\tau}|$ significantly modifies the probability, whereas the changes due to $a_{\mu\mu}$ and $a_{\tau\tau}$ are negligibly small. 
In all cases, the positive or negative values  of the LIV parameters are responsible for  the decrease or enhancement of the oscillation probabilities. In the middle (right) panel of  Fig. \ref{pro-d}, we show the change (relative change) in oscillation probability due to the effect of LIV parameters.

\section{Sensitivity Limits on the LIV parameters}
\begin{figure}
\includegraphics[scale=0.34]{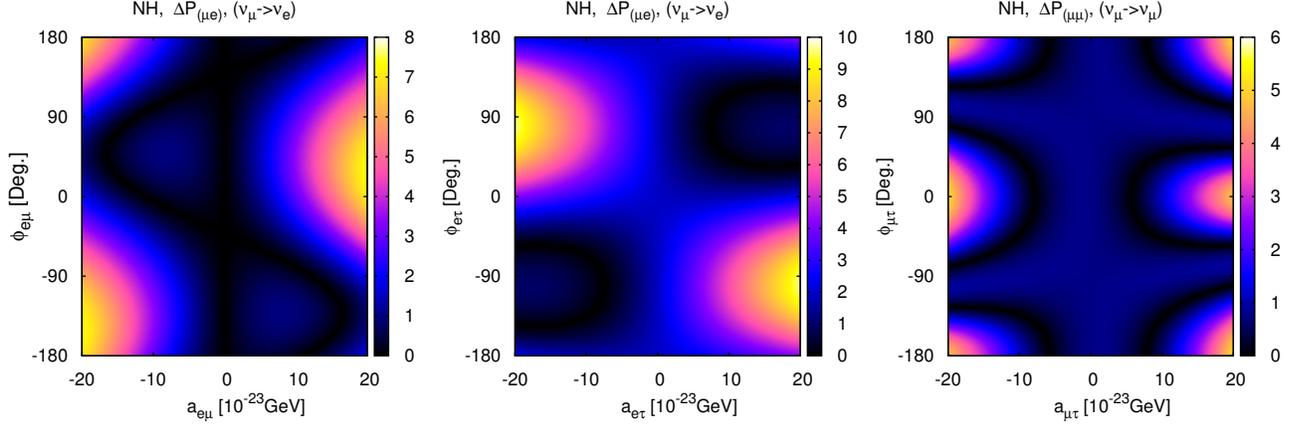}
\caption{Representation of $\Delta P_{\mu e}$ and $\Delta P_{\mu \mu}$ in  $a_{\alpha\beta}-\phi_{\alpha\beta}$ LIV parameter space for NO$\nu$A experiment. The left (middle) panel is for the sensitivities of $\Delta P_{\mu  e}$ in $a_{e\mu}-\phi_{e\mu}$ ($a_{e\tau}-\phi_{e\tau}$) plane and right panel is for  $\Delta P_{\mu\mu}$ in the $a_{\mu\tau}-\phi_{\mu\tau}$ plane. The color bars in right side of each plot represent the relative change of the $\Delta P_{\alpha\beta}$ in the corresponding plane.}
\label{oscillo}
\end{figure}
In this section, we analyse the potential of   T2K, NO$\nu$A, and the synergy of T2K and NO$\nu$A to constrain the LIV parameters. From Eqns. (16) and (18) or from  Fig. \ref{pro-nd} and Fig. \ref{pro-d}, it can be seen that the LIV parameters $|a_{e\mu}|$ and $|a_{e\tau}|$ along with LIV phases $\phi_{e\mu}$ and $\phi_{e\tau}$ play major role in appearance channel ($\nu_\mu \to \nu_e$), whereas $|a_{\mu\tau}|$ and $\phi_{\mu\tau}$   influence the survival  channel ($\nu_\mu \to \nu_\mu$). In order to see their sensitivities at probability level, we define two quantities, $\Delta P_{\mu e} =\frac{|P_{\mu e}^{\rm LIV}-P_{\mu e}^{\rm SM}|}{P_{\mu e}^{\rm SM}}$ and  $\Delta P_{\mu \mu} =\frac{|P_{\mu \mu}^{\rm LIV}-P_{\mu \mu}^{\rm SM}|}{P_{\mu \mu}^{\rm SM}}$, which provide the information about  the relative change  in probability due to the presence of  LIV term from the standard case. We  evaluate their values for various  LIV parameters and display them  in $a_{\alpha \beta}- \phi_{\alpha \beta}$ plane in  Fig. \ref{oscillo}.  From the left panel of the  figure, one can see that the observable $\Delta P_{\mu e}$ has maximum value at the  yellow region, for $\phi_{e\mu} \approx 45^{\circ}$, if  $a_{e\mu}$ is positive, whereas for negative value of $a_{e\mu}$, $\Delta P_{\mu e}$ is maximum for $\phi_{e\mu} \approx -135^{\circ}$. This nature of  $\Delta P_{\mu e}$  can be easily understood from  Eq. (16), as the appearance probability depends on sine and cosine functions of $\phi_{e\mu}$. However, the nature of $\Delta P_{\mu e}$ for $e\tau$ sector is quite different from that of $e\mu$ sector,  even-though the appearance probability depends upon sine and cosine functions of $\phi_{e\tau}$. This is due to the opposite sign on $|a_{e\mu}|$ and $|a_{e\tau}|$ dependent terms in oscillation probability. As the LIV parameter $|a_{\mu\tau}|$ mainly appears  on the survival channel, we  calculate $\Delta P_{\mu \mu}$ which has cosine dependence on $\phi_{\mu\tau}$ and display it in the right panel of the figure.

\begin{figure}
\begin{center}
\includegraphics[scale=0.32]{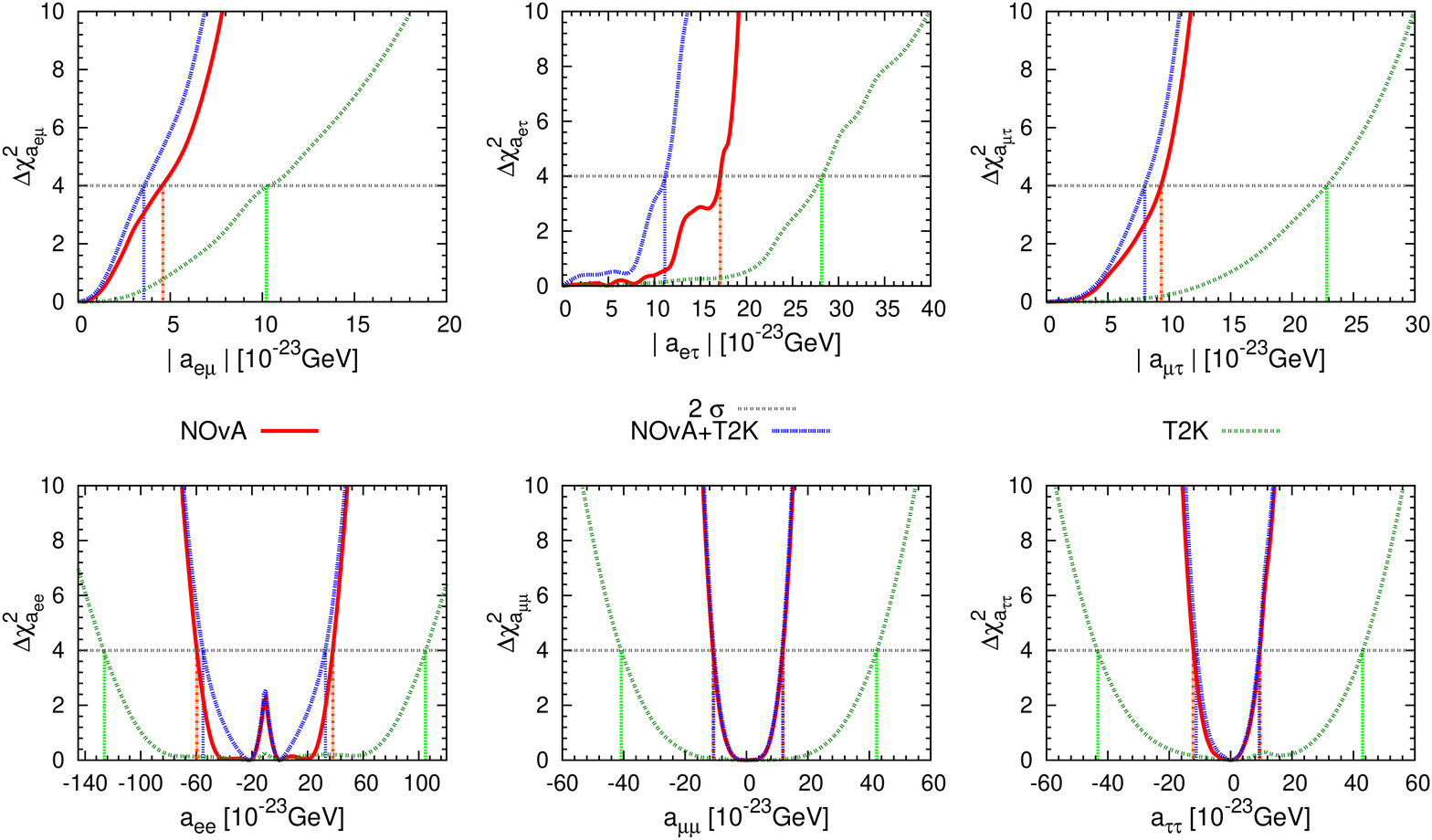}\\
\caption{The sensitivities on LIV parameters from  NO$\nu$A and T2K experiments.}
\label{sens}
\end{center}
\end{figure}

Next, we analyze the potential of  T2K, NO$\nu$A,  and the synergy of T2K  and NO$\nu$A  to constrain the various LIV parameters, which are shown  in Fig. \ref{sens}. In order to obtain these values, we compare the true event spectra which are generated in the standard three flavor oscillation paradigm with the test event spectra which are simulated by including one LIV parameter at a time and  show the marginalized sensitivities as a function of the LIV parameters,  $|a_{\alpha \beta}|$.  The values of $\Delta\chi^2_{\alpha\beta}$ are evaluated using the standard rules as described in GLoBES and the details are presented in the Appendix.  From the figure, we can see that the sensitivities on LIV parameters obtained from T2K are much weaker than NO$\nu$A and the synergy of T2K  and NO$\nu$A can improve the sensitivities on these  parameters. For a direct comparison,  we give the sensitivity limits on each LIV parameter (in GeV) at 2$\sigma$ C.L. in Table II. 
\begin{table}[]
\begin{tabular}{|c|c|c|c|}
\hline
\multirow{2}{*}{LIV parameter} & \multicolumn{3}{c|}{Sensitivity limit on LIV parameter} \\ \cline{2-4} 
                     & T2K  & NO$\nu$A  & T2K+NO$\nu$A \\ \hline
      $|a_{e\mu}|$     &  $<1.02\times 10^{-22}$    & $< 0.46\times 10^{-22}$        &  $<0.36\times 10^{-22}$  \\ \hline
      $|a_{e\tau}|$    &   $<2.82\times 10^{-22}$   & $< 1.71\times 10^{-22}$       &  $<1.08\times 10^{-22}$  \\ \hline
      $|a_{\mu\tau}|$  &  $<2.28\times 10^{-22}$    & $< 0.93\times 10^{-22}$       &   $<0.8\times 10^{-22}$ \\ \hline
      $a_{ee}$         &   $[-12.62:10.47]\times 10^{-22}$   & $[-5.97:3.82 ] \times 10^{-22}$ & $[-5.52:3.29]\times 10^{-22}$   \\ \hline
      $a_{\mu\mu}$     &   $[-4.09:4.24]\times 10^{-22}$     & $[-1.09:1.19]\times 10^{-22}$   & $[-1.07:1.18]\times 10^{-22}$    \\ \hline
     $a_{\tau\tau}$    &   $[-4.33:4.3]\times 10^{-22}$   & $[-1.22:0.96]\times 10^{-22}$   &  $[-1.12:0.93]\times 10^{-22}$    \\ \hline

\end{tabular}
\label{bounds}
\caption{The sensitivity limits on each LIV parameters (in GeV) at 2$\sigma$ C.L. from T2K, NO$\nu$A, and synergy between T2K and  NO$\nu$A. }
\end{table}
All these limits are slightly weaker than the bounds obtained from Super-Kamiokande Collaboration (\ref{sk}). 
\section{Effect of LIV on various sensitivities of NO$\nu$A}
In this section, we discuss the effect of LIV on the sensitivities of long-baseline experiment to  determine neutrino mass ordering and CP-violation by taking NO$\nu$A  as a case of study.  In addition to this, we also present the correlations between the LIV parameters and the standard oscillation parameters $\theta_{23}$ and $\delta_{CP}$.
\subsection{CP violation discovery potential}
\begin{figure}
\hspace*{-1cm}
\includegraphics[scale=0.28]{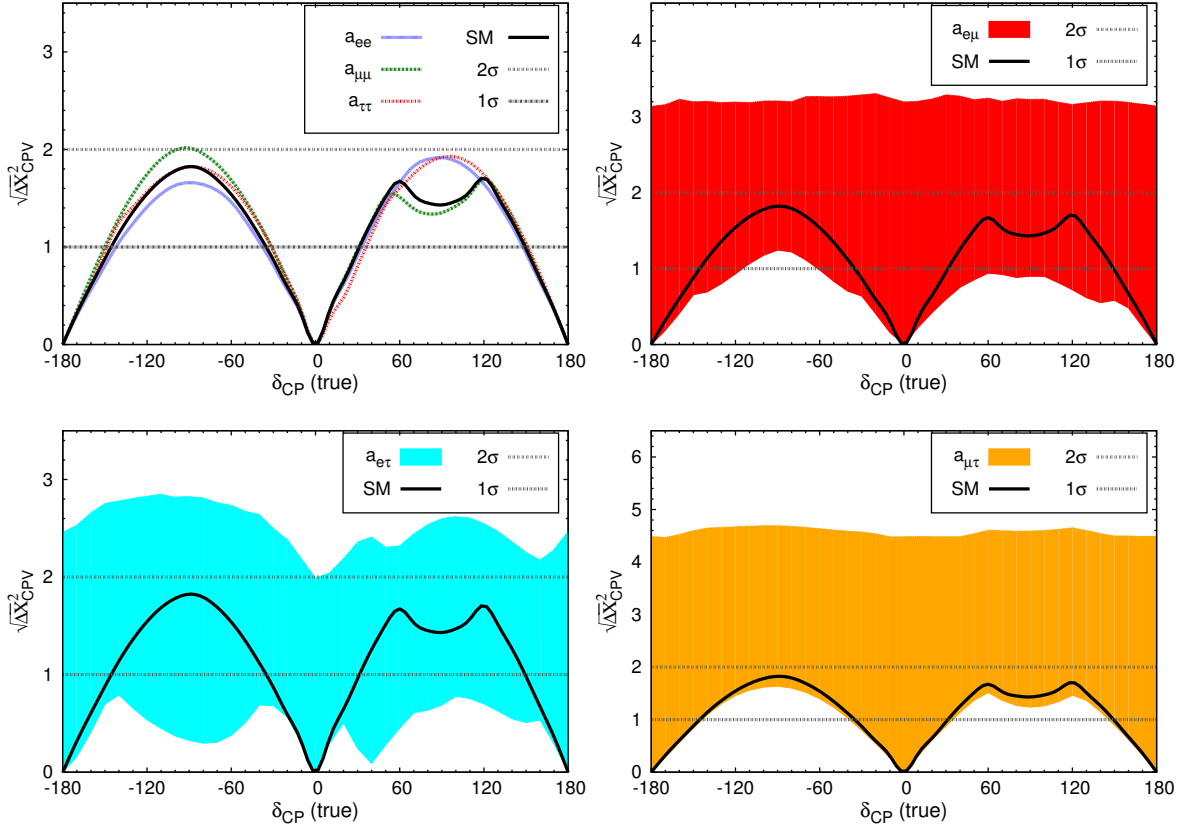}
\caption{CP Violation sensitivity as a function of true values of $\delta_{CP}$ for NO$\nu$A experiment. Standard case is represented by black curve in each plot. The top-left panel is for diagonal Lorentz violating parameters and non-diagonal LIV parameters in $e\mu$, $e\tau$ and $\mu\tau$ sectors shown in top-right, bottom-left and bottom-right panels respectively.}
\label{cpv}
\end{figure}

It is well known that the determination of the CP violating phase $\delta_{CP}$ is one of the most challenging issues in neutrino physics today. CP violation in the leptonic sector may provide the  key ingredient  to explain the observed baryon asymmetry of the Universe through  leptogenesis. In this section, we discuss how the  CP violation sensitivity of NO$\nu$A experiment gets affected due to impact of LIV parameters. 
Fig. \ref{cpv}  shows the significance with which CP violation, i.e. $\delta_{CP} \neq 0, \pm \pi$  can be determined for different true values of $\delta_{CP}$.  For the calculation of sensitivities, we have used the oscillation parameters as mentioned in  Table \ref{P.table}. Also, the amplitude of all the diagonal LIV parameters considered as $1\times 10^{-22}$ GeV and non-diagonal elements as $2\times 10^{-23}$ GeV. The expression for the test statistics $\Delta\chi^2_{CPV}$, which quantifies the CP violation sensitivity is provided in the Appendix. We consider here the true hierarchy as normal, true parameters as given in  Table \ref{P.table},  and vary the true value for $\delta_{CP}$ in the allowed range $[-\pi,\pi]$. Also the possibility of exclusion of CP conserving phases has been shown by taking the test spectrum $\delta_{CP}$ value as 0, $\pm \pi$. This exclusion sensitivity is obtained by calculating the minimum $\Delta\chi^2_{\rm min}$ after doing   marginalization  over both hierarchies NH and IH,  as well as $\Delta m^2_{31}$ and  $\sin^2\theta_{23}$ in their $3\sigma$ ranges. The CPV sensitivity for standard case and in presence of diagonal LIV parameters is shown in the top left panel of Fig. \ref{cpv}. The black curve depicts the standard case, and for diagonal elements  $a_{ee},~a_{\mu\mu}$ and $a_{\tau\tau}$, the corresponding plots are displayed by blue, green and red respectively. Further, we show the sensitivity in presence of non-diagonal LIV parameters in $e\mu$, $e\tau$, and $\mu\tau$ sectors respectively in the top right, bottom left, and bottom right panels of the same figure.   As the extra phases of the non-diagonal parameters can affect the CPV sensitivity,  we calculate the value of $\Delta\chi^2_{\rm min}$ for a particular value of $\delta_{CP}$ by varying the phase $\phi_{\alpha \beta}$  in its allowed range $[-\pi, \pi]$, which results in a band structure. It can be seen from figure that LIV can significantly affect the CPV discovery potential of the NO$\nu$A experiment. All the three non-diagonal LIV  parameters have significant impact on CPV sensitivity. It can be  seen from the figure that CPV sensitivity spans on both sides of standard case in presence of non-diagonal LIV parameters. Although there is a possibility that the  sensitivity can be deteriorated in presence of LIV for some particular true value of the phase of the non-diagonal parameter ($\phi_{\alpha \beta})$, for most of the case  the  CP violation sensitivity is significantly get enhanced.  Moreover, one can expect some sensitivity where there is less or no such significance for $\delta_{CP}$ regions in standard case. Further, the parameters $a_{e\mu}$ and $a_{e\tau}$ have comparatively large effect on the sensitivity with respect to to $a_{\mu\tau}$.   Similar observation can also be  found by considering inverted hierarchy.
\begin{center}
\begin{figure}
\includegraphics[scale=0.45]{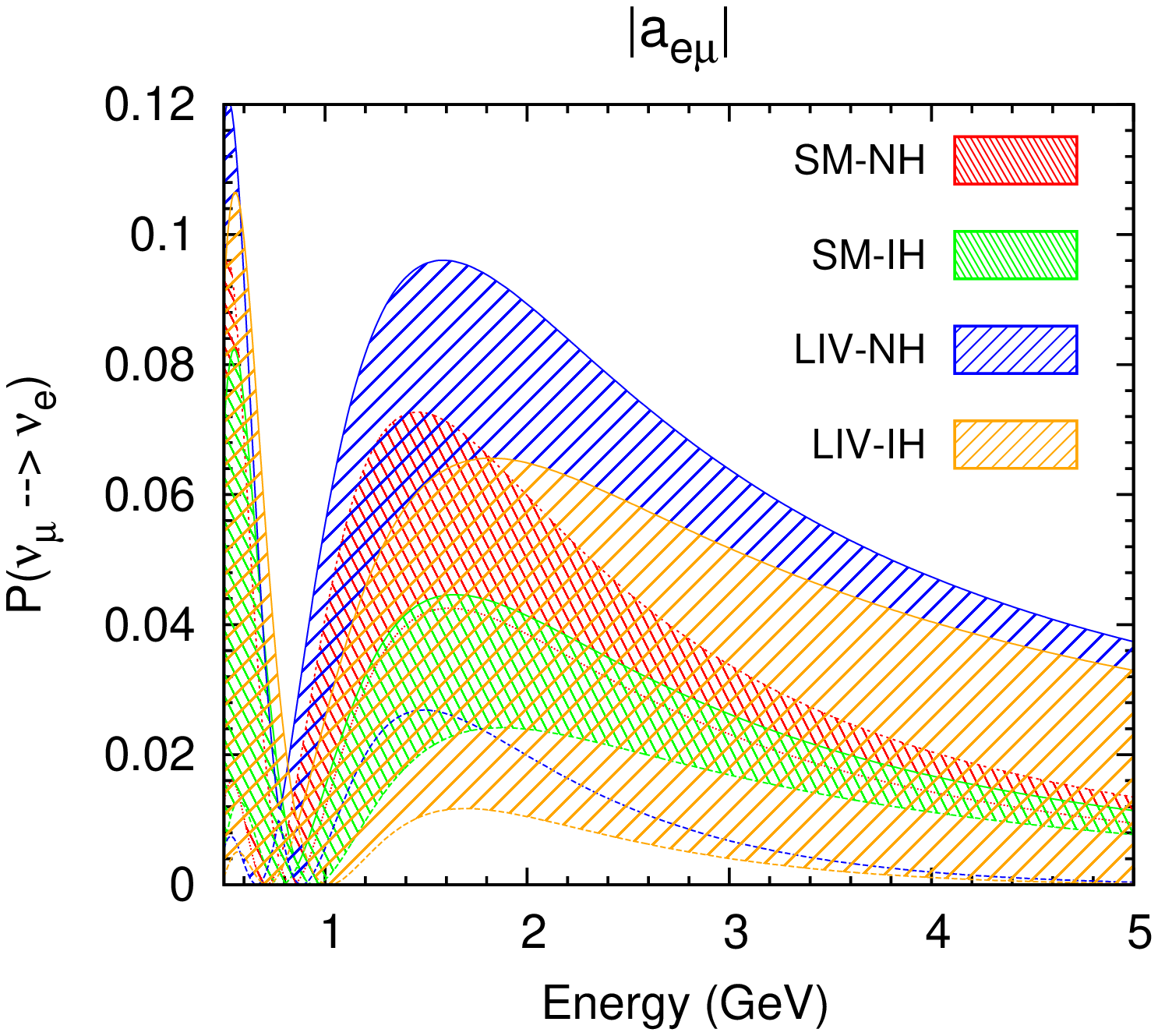}
\includegraphics[scale=0.45]{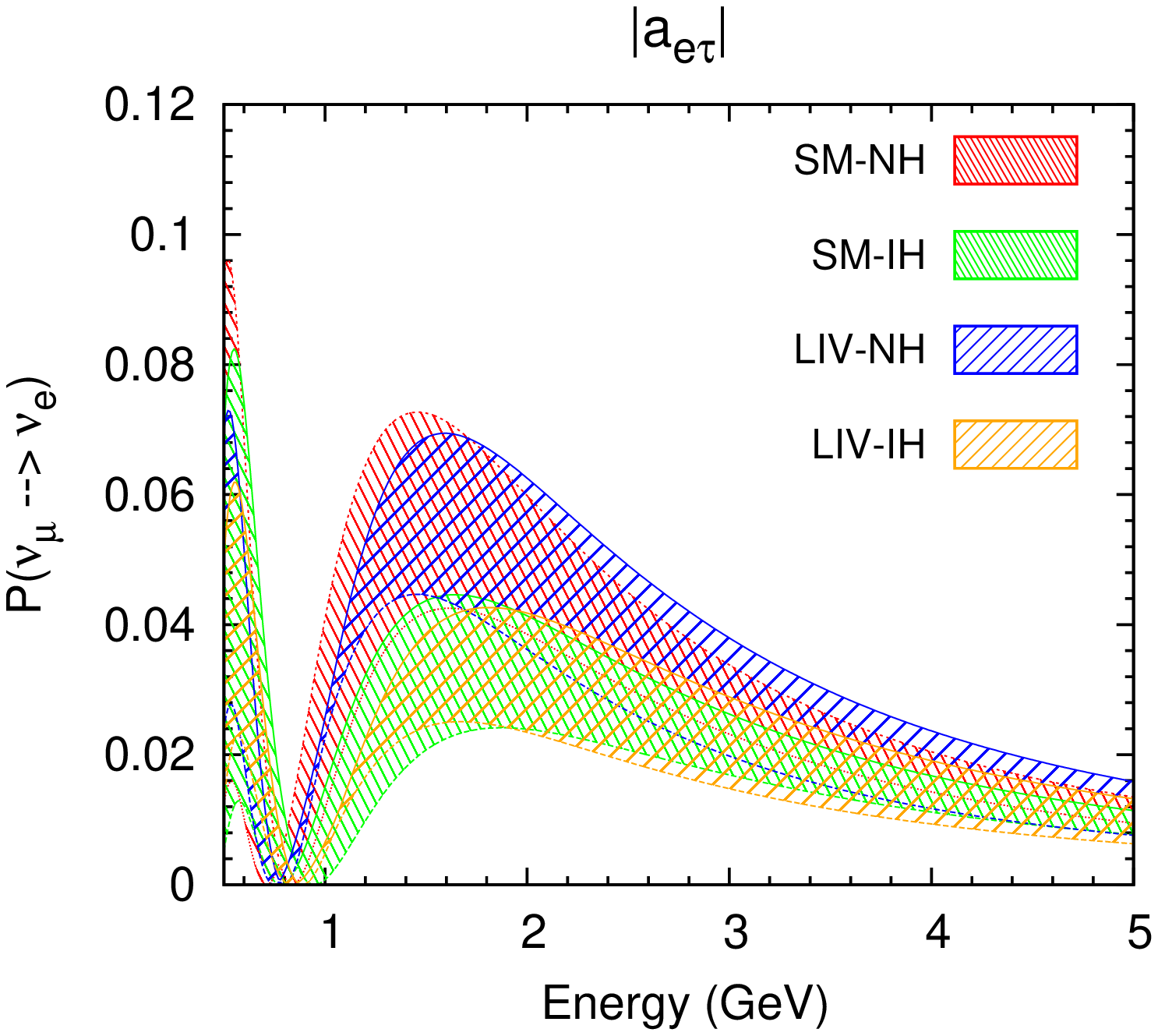}\\
\includegraphics[scale=0.45]{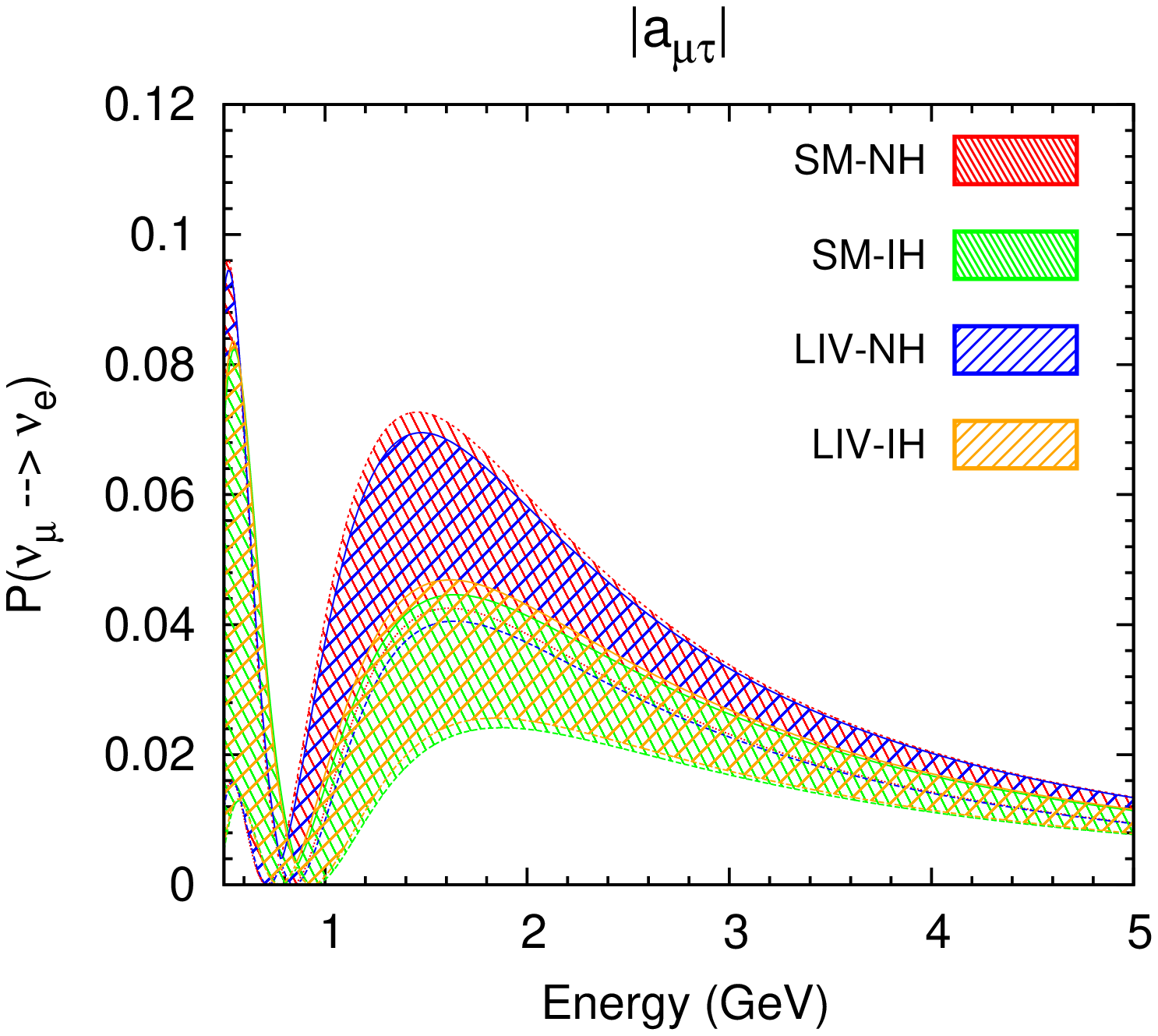}
\includegraphics[scale=0.45]{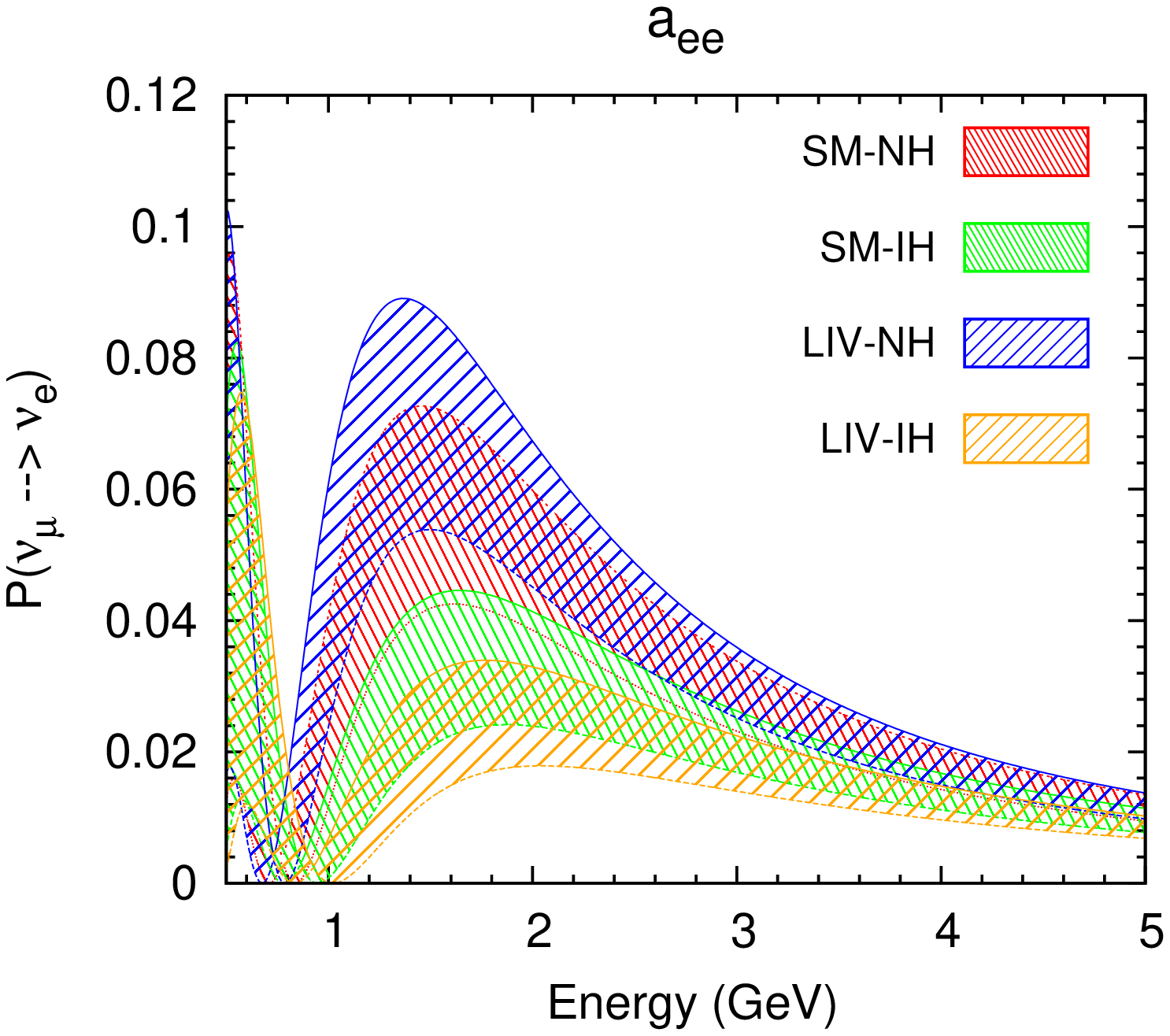}
\caption{The oscillation probability for NO$\nu$A experiment as a function of energy in presence of non-diagonal LIV parameters $|a_{e\mu}|, |a_{e\tau}|$ and $|a_{\mu\tau}|$ are shown in top-let, top-right and bottom-left panels respectively. The effect of diagonal LIV parameter $a_{ee}$ shown in bottom-right panel.} 
\label{pro-band-nd}
\end{figure}
\end{center}
\subsection{MH Sensitivity}
Mass hierarchy determination is one of the main objectives of the long baseline experiments. It is determined by considering true hierarchy as NH (IH) and comparing it with the test hierarchy, assumed to be  opposite to the true case, i.e.,  IH (NH).  Fig. \ref{pro-band-nd} shows the effect of LIV parameters on MH sensitivity at oscillation probability level. We obtain the bands by varying the $\delta_{CP}$ within its allowed range $[-\pi,\pi]$ and considering the other parameters   as given in the Table \ref{P.table},  and the amplitude of all the non-diagonal LIV elements as $2\times 10^{-23}$ GeV and diagonal LIV elements as $1\times 10^{-22}$ GeV. The red (green) band in the figure is for NH (IH) case with standard matter effect. There is some overlapped region between the two bands for some values of $\delta_{CP}$, where determination of neutrino mass ordering is difficult. The blue and orange bands represent the NH and IH case in presence of the LIV parameters respectively. It can be seen that the parameter $a_{e\mu}$ and $a_{ee}$ have significant effect on the appearance probability energy spectrum compared to other two parameters. The two bands NH and IH shifted to higher values of probability and have more overlapped regions  in presence of $a_{e\mu}$. The presence of $a_{ee}$ shifted the NH band to higher values and IH band shifted to lower values of probabilities compared to standard case. Whereas the effects of $a_{e\tau}$ and $a_{\mu\tau}$ are negligibly small. 

\begin{figure}[htbp]
\includegraphics[scale=0.5]{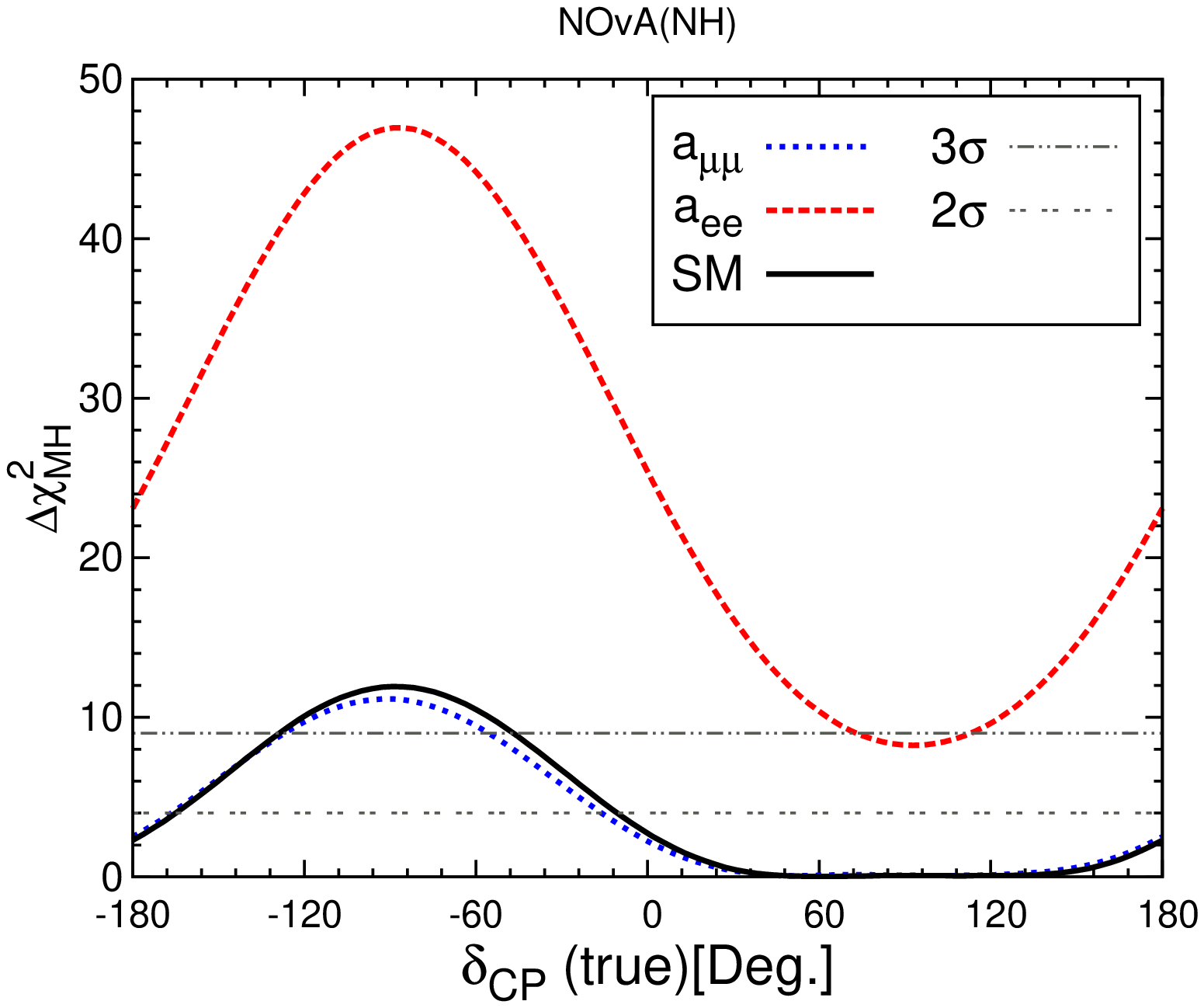}
\includegraphics[scale=0.5]{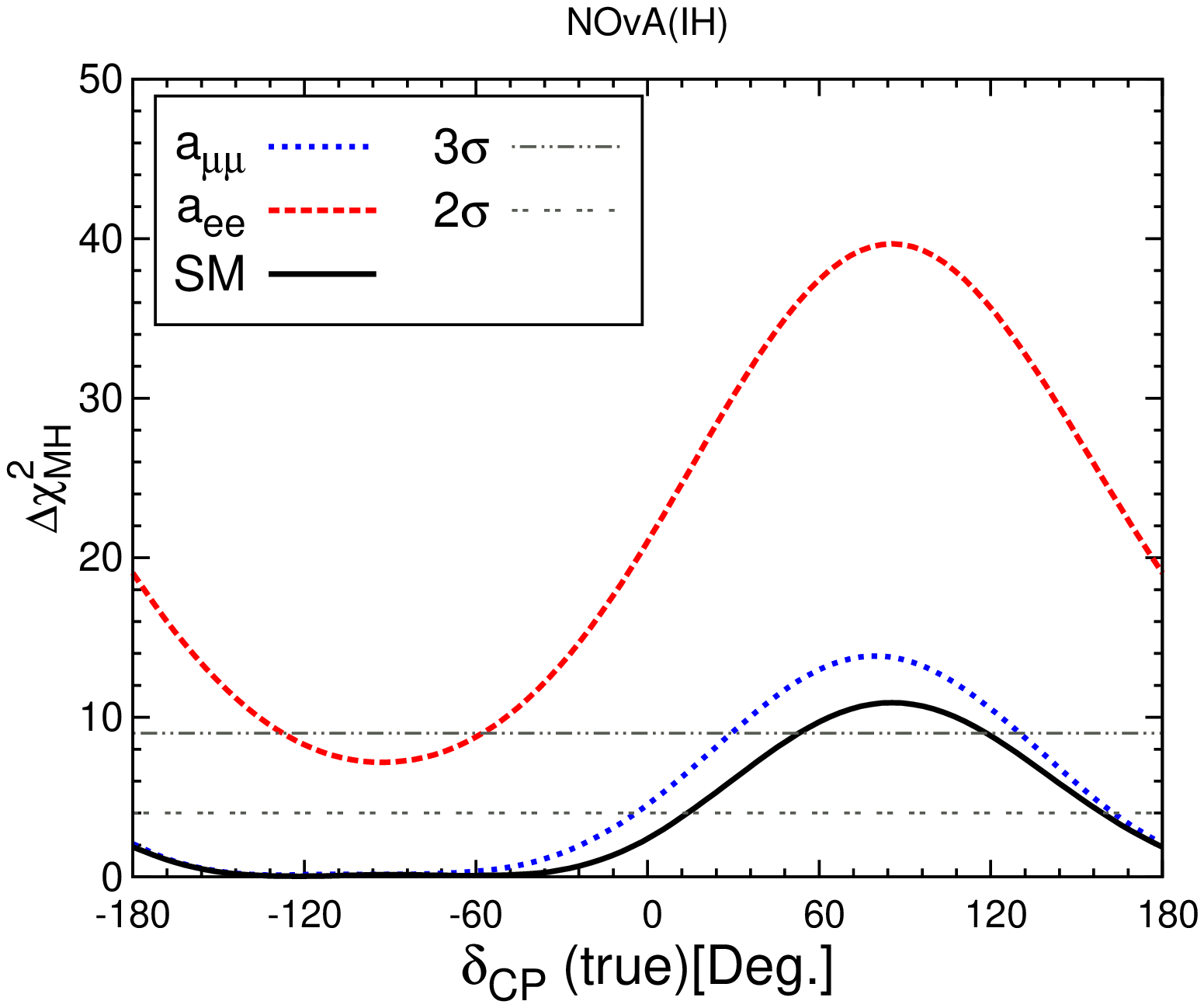}
\caption{Mass hierarchy sensitivity as function of $\delta_{CP}$ for NO$\nu$A experiment. Left (right) panel is for NH (IH) as true value. Black curve represents the standard matter effect case without any LIV parameter. Red and blue dotted  curves represent the sensitivity in the presence of diagonal parameters $a_{ee}$, and $a_{\mu\mu}$  respectively.}
\label{mh-sen}
\end{figure}

 Next, we calculate the $\Delta\chi^2_{MH}$ by comparing true event and test event spectra which are generated for the oscillation parameters in the Table \ref{P.table} for each true value of  $\delta_{CP}$. In order to get the minimum deviation or $\Delta\chi^2_{\rm min}$, we do marginalization over $\delta_{CP},~\theta_{23}$ and $\Delta m^2_{31}$ in their allowed regions. In Fig. \ref{mh-sen}, we show the mass hierarchy sensitivity of NO$\nu$A experiment for standard paradigm  and  in presence of diagonal LIV parameter.  The left (right) panel of the figure corresponds to the MH sensitivity for true NH (IH). It can be seen from the figure that for standard matter effect case (black curve), the test hierarchy can be ruled out in upper half plane (UHP) ($0<\delta_{CP}<\pi$) and lower half plane (LHP) ($-\pi<\delta_{CP}<0$) for true NH and IH respectively above 2$\sigma$ C.L.. The other half plane is unfavourable for mass hierarchy determination. The parameter $a_{ee}$ is found to give significant enhancement from the standard case compared to  $a_{\mu\mu}$.  

It should also be emphasized that mass hierarchy can be measured precisely above 3$\sigma$ C.L. for most of the $\delta_{CP}$ region in presence of $a_{ee}$ for true value in both NH and IH. 

\begin{center}
\begin{figure}[htbp]
\includegraphics[scale=1]{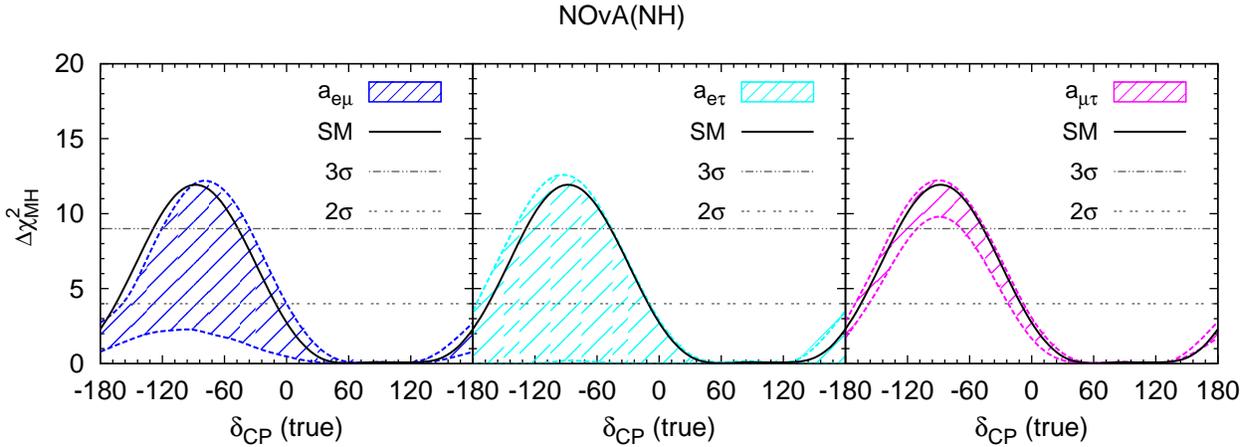}
\caption{Mass hierarchy sensitivity as a function of $\delta_{CP}$ for NO$\nu$A experiment in presence of $a_{\alpha \beta}$. Black curve represents the standard matter effect case without any LIV parameter. Left, middle and right panels represent the sensitivity in presence of non-diagonal parameters $a_{e\mu},~a_{e\tau}$ and $a_{\mu\tau}$ respectively.}
\label{mh-sen-2}
\end{figure}
\end{center}

The MH sensitivity in presence non-diagonal Lorentz violating parameters $a_{\alpha \beta}$ is shown in Fig. \ref{mh-sen-2}. As the non-diagonal LIV parameters introduce new phases, we do  marginalization over new  phases in their allowed range, i.e., $[-\pi,\pi]$ while obtaining the MH sensitivity. In all the three cases, the MH sensitivity expands around the MH sensitivity in the standard three flavor framework. 
 From the figure, it can be seen that the non-diagonal LIV parameters significantly affect the sensitivity which  crucially depends on the value of new phase. Similar analysis can be studied considering IH as the true hierarchy. 

\subsection{Correlations between LIV parameters with $\delta_{CP}$ and $\theta_{23}$}
\begin{figure}
\includegraphics[scale=0.33]{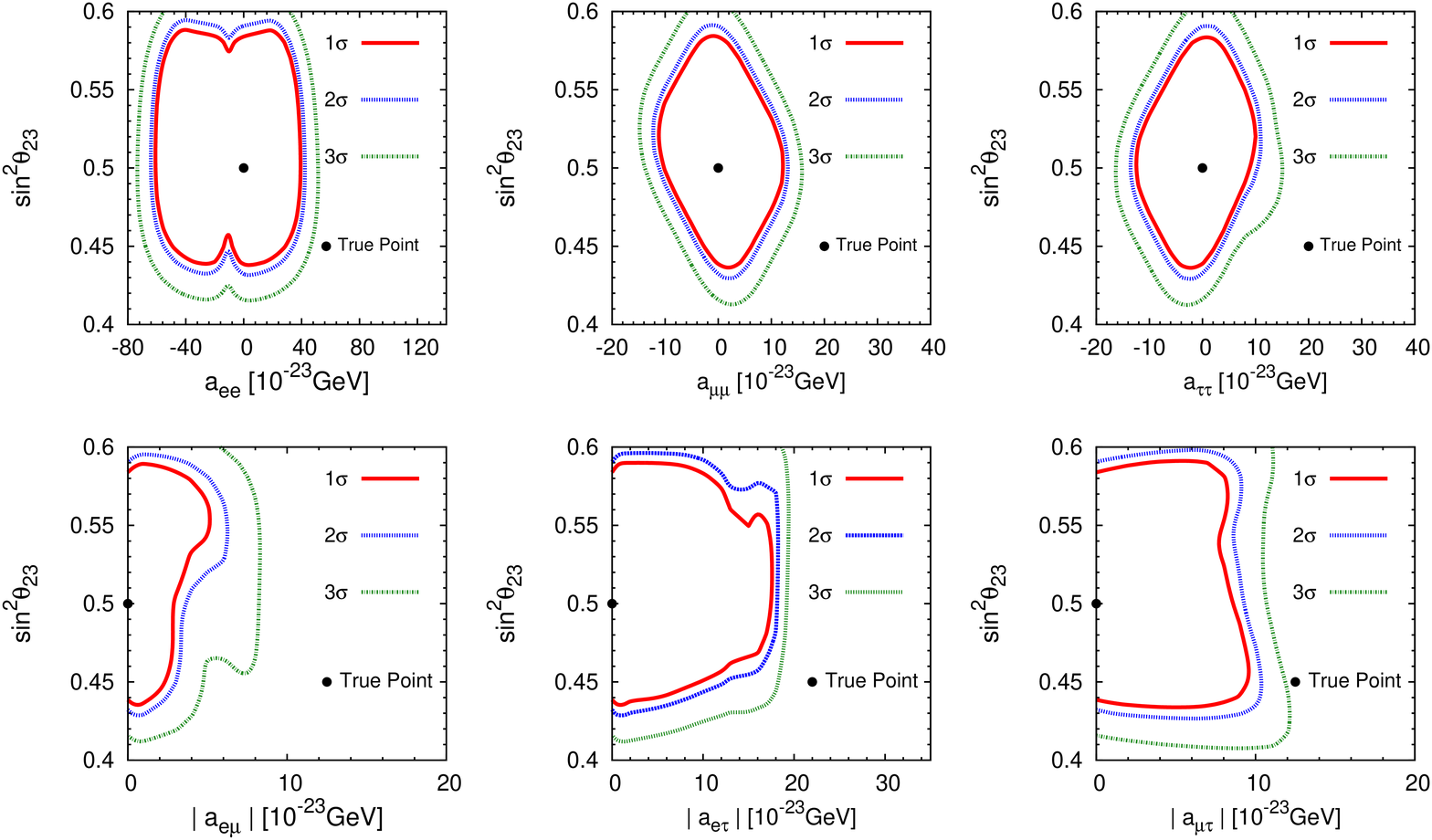}
\caption{ Correlation between LIV parameters and $\theta_{23}$ in $|a_{\alpha\beta}|-\sin^2\theta_{23}$ plane  at 1$\sigma$,~2$\sigma$ and 3$\sigma$ C.L. for NO$\nu$A experiment.}
\label{th23liv}
\end{figure}

\begin{figure}
\includegraphics[scale=0.33]{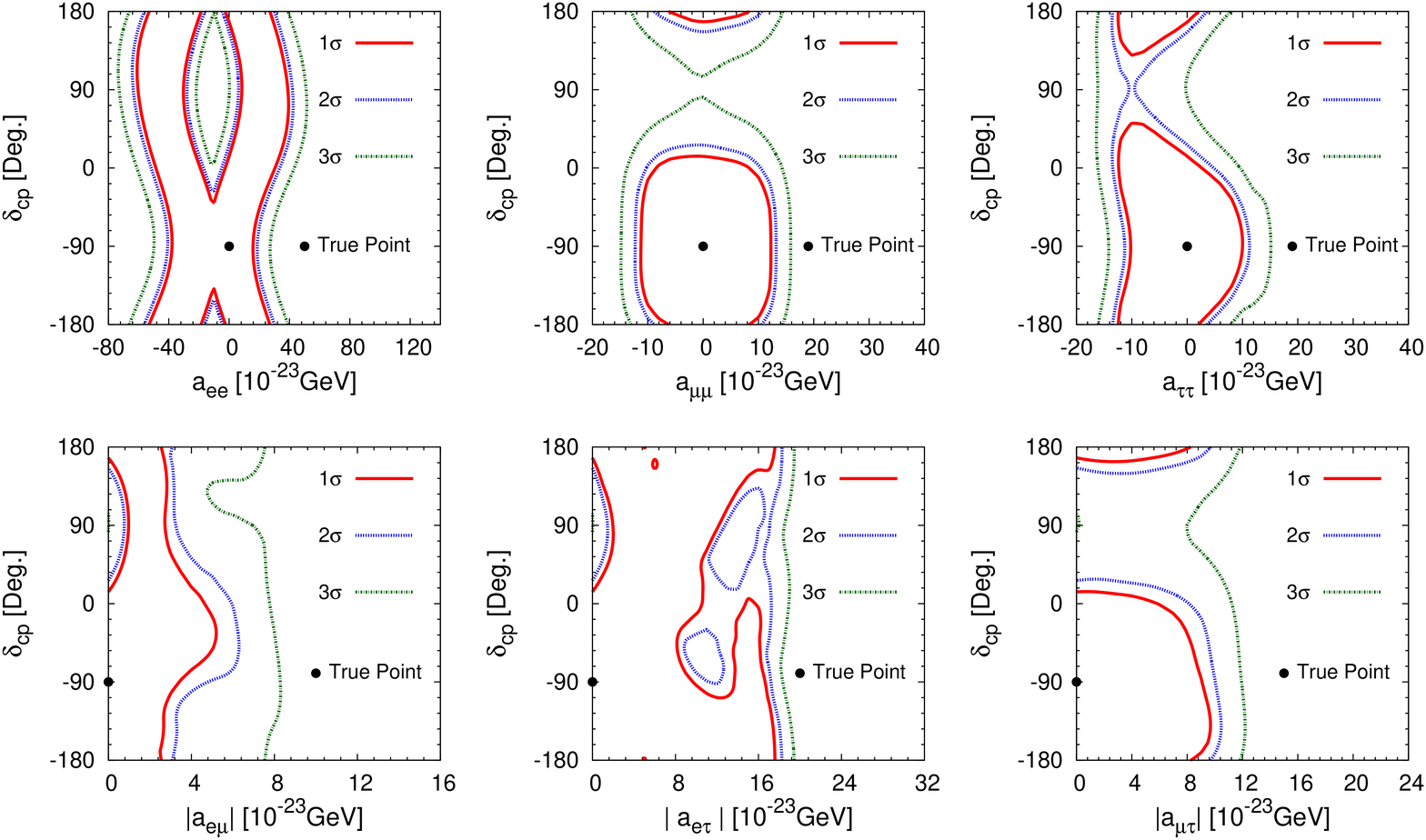}
\caption{Correlation between LIV parameters and $\delta_{CP}$ in $|a_{\alpha\beta}|-\delta_{CP}$ plane at 1$\sigma$,~2$\sigma$ and 3$\sigma$ C.L. for NO$\nu$A experiment.}
\label{deltaliv}
\end{figure}

In this section, we show the correlation between the LIV parameters and the standard oscillation parameters $\theta_{23}$ and $\delta_{CP}$ in $|a_{\alpha\beta}|-\theta_{23}$ and $|a_{\alpha\beta}|-\delta_{CP}$  planes.  Fig. \ref{th23liv} (\ref{deltaliv}) shows the  correlation for  $a_{ee}$, $a_{\mu\mu}$, $a_{\tau\tau}$, $|a_{e\mu}|$, $|a_{e\tau}|$, $|a_{\mu\tau}|$ and $\theta_{23}$ ($\delta_{CP}$), at 1$\sigma$, 2$\sigma$, 3$\sigma$ C.L. in two dimensional plane. In both figures upper (lower) panel is for $a_{ee}$, $a_{\mu\mu}$ and $a_{\tau\tau}$ ($|a_{e\mu}|$, $|a_{e\tau}|$, $|a_{\mu\tau}|$). In order to obtain these correlations, we set the true value of LIV parameters  to zero and the standard oscillation parameters as given in Table \ref{P.table}.  
Further, we do marginalization over $\sin ^2\theta_{23},\delta_{CP},$ and $\Delta m^2_{31}$ for both hierarchies. In the case of non-diagonal LIV parameters,  $|a_{e\mu}|,|a_{e\tau}|,|a_{\mu\tau}|$, we  also do  marginalization over the additional phase $\phi_{\alpha\beta}$. From the plots it can be noticed that precise determination of $\theta_{23}$ will provide useful information about the possible interplay of LIV physics.


\medskip
 
 \section{Summary and Conclusion}
It is well known that,  neutrino oscillation physics has entered a precision era, and the currently running accelerator based long-baseline experiment  NO$\nu$A is  expected to  shed light on the current unknown parameters in the standard oscillation framework, such as the mass ordering as well as the leptonic CP phase $\delta_{CP}$. However, the possible interplay of  potential new physics scenarios can hinder the clean determination of these parameters.  Lorentz invariance is one of the fundamental properties of space time in the standard version of relativity.  Nevertheless,  the possibility of small violation of this fundamental symmetry has been explored in various extensions of the SM in recent times and a   variety of possible experiments  for the search of such signals have been proposed over the years.  In this context, the study of neutrino properties can also provide a suitable testing ground to look for the effects of LIV parameters as 
neutrino phenomenology is extremely  rich and spans over a very wide range of energies. In this work, we have studied in detail the impact of Lorentz Invariance violating parameters on the currently running long-baseline experiments T2K and NO$\nu$A and our findings are summarized below.
\begin{itemize}
\item Considering the effect of only one LIV parameter at a time, we have obtained the  sensitivity limits on these  parameters for the currently running long baseline experiments T2K and NO$\nu$A. We found that the limits obtained from T2K are much weaker than that of NO$\nu$A  and the  synergy of T2K and NO$\nu$A  can significantly improve these sensitivities. 
\item 
 We have  also explored  the  phenomenological  consequences  introduced  in the neutrino oscillation physics due to the presence of Lorentz-Invariance violation on the sensitivity studies of  long-baseline experiments by considering NO$\nu$A as a case  study. We mainly focused on how the oscillation probabilities, which govern the neutrino flavor transitions, get modified  in presence  of  different LIV parameters. 
  In particular, we have considered the impact of the LIV parameters $|a_{e \mu}|,~|a_{e \tau}|$,  $|a_{\mu \tau}|$, $a_{ee}$, $a_{\mu\mu}$  and  $a_{\tau \tau}$.  We found that the parameters 
  $|a_{e \mu}|$,  $|a_{e \tau}|$ and $a_{e e}$ significantly affect the $\nu_\mu \to \nu_e$ transition  probability $P_{\mu e}$, while the effect of  $|a_{ \mu \tau}|$,   $a_{\mu\mu}$, $a_{\tau \tau}$ on the survival probability $P_{\mu \mu}$ is minimal.   We also found that 
   $|a_{e \mu}|$  creates a distortion on the appearance probability.  
\item   We  further investigated the impact of LIV parameters on the determination of mass hierarchy and CP violation discovery potential and found that the presence of LIV parameters significantly affect these sensitivities.  In fact, the mass hierarchy sensitivity and CPV sensitivity are enhanced or deteriorated significantly in presence of LIV parameters as these sensitivities  crucially depend  on the new CP-violating phase of these parameters. 

\item We also obtained the correlation plots between $\sin^2 \theta_{23}$ and $|a_{\alpha \beta}|$ as well as between $\delta_{CP}$ and $|a_{\alpha \beta}|$. From these confidence regions, it can be ascertained that it is possible to obtain the limits on the LIV parameters once $\sin^2 \theta_{23}$ is precisely determined.
 \end{itemize}
 In conclusion, we found that T2K and NO$\nu$A have the potential to explore the new physics associated with Lorentz invariance violation and can provide constraints on these parameters.

\section*{Appendix: Details of $\chi^2$ analysis}\label{chi}

In our analysis, we have performed the $\chi^2$ analysis by comparing true (observed) event spectra  $N_i^\textrm{true}$ with test  (predicted) event spectra $N_i^\textrm{test}$,  and its general form is given by
 \begin{equation}
\chi^2_{\rm stat} (\vec{p}_\textrm{true},\vec{p}_\textrm{test})= -\sum_{i\in\textrm{bins}} 2\Big[N_i^\textrm{test} - N_i^\textrm{true}-N_i^\textrm{true} \ln\left(\frac{N_i^\textrm{test}}{N_i^\textrm{true}}\right) \Big],
\end{equation}
where $\vec{p}$ is the array of standard neutrino oscillation parameters. However, for numerical calculation of  $\chi^2$, we also include the systematic errors using pull method. This is  usually done with the help of nuisance systematic parameters as discussed in the GLoBES manual.  In presence of systematics, the predicted event spectra modify as $N_i^\textrm{test} \to N_i^{'~\textrm{test}}= N_i^\textrm{test} (1+\sum_{j=1}^n \pi_i^j\xi_j^2)$, where $\pi_i^j$ is the systematic error associated with signals and backgrounds and $\xi_j$ is the pull. Therefore, the Poissonian $\chi^2$ becomes
\begin{equation}
\chi^2 (\vec{p}_\textrm{true},\vec{p}_\textrm{test},\vec{\xi})= -\underset{\vec{\xi_j}}{\mathrm{ min}} \sum_{i\in\textrm{bins}} 2\Big[N_i^{'~\textrm{test}} - N_i^\textrm{true}-N_i^\textrm{true} \ln\left(\frac{N_i^{'~\textrm{test}}}{N_i^\textrm{true}}\right) \Big]+ \sum_{j=1}^{n} \xi_j^2.
\end{equation}
Suppose $\vec{q}$ is the oscillation parameter in presence of Lorentz invariance violating parameters. Then the sensitivity of LIV  parameter $a_{\alpha \beta}$ can be evaluated as
\begin{equation}
 \Delta\chi^2(a_{\alpha \beta}^\text{test}) = \chi^2_{\rm SO}- \chi^2_{\rm LIV}\;,\end{equation}
where  $\chi^2_{\rm SO}= \chi^2(\vec{p}_\textrm{true},\vec{p}_\textrm{test})$,
$\chi^2_{\rm LIV}= \chi^2(\vec{p}_\textrm{true},\vec{q}_\textrm{test})$. We obtain minimum $\Delta\chi^2(a_{\alpha \beta}^\text{test})$ by doing marginalization over $\sin^2\theta_{23}$, $\delta m_{31}^2$, and $\delta_{\mathrm{CP}}$. Further, the sensitivities of current unknowns in neutrino oscillation is given by
\begin{itemize}
\item CPV sensitivity: 
\begin{equation}
\Delta\chi^2_\text{CPV} (\delta^\text{true}_{CP}) = \text{min}[\chi^2(\delta^\text{true}_{CP},\delta^\text{test}_{CP}=0), \chi^2 (\delta^\text{true}_{CP},\delta^\text{test}_{CP}=\pi)].
\end{equation}

\item MH sensitivity: 
\begin{eqnarray}
\Delta\chi^2_\text{MH} &=& \chi^2_\text{NH} -\chi^2_\text{IH}~~~~(\text{for true normal ordering}),\\
\Delta\chi^2_\text{MH} &=& \chi^2_\text{IH} -\chi^2_\text{NH}~~~~(\text{for true inverted ordering}).
\end{eqnarray}
\end{itemize}
Further, we obtain minimum  $\chi^2_\text{MH}$ by doing marginalization over the oscillation parameters  $\sin^2 \theta_{23}$, $\Delta m^2_{31}$, and $\delta_{CP}$ in the range [0.4:0.6], [2.36:2.64]$\times 10^{-3}$ eV$^2$ and [$-180^\circ$,180$^\circ$] respectively, and for obtaining minimum  $\chi^2_\text{CPV}$  marginalization is done over the oscillation parameters  $\sin^2 \theta_{23}$ and $\Delta m^2_{31}$.
While including the non-diagonal Lorentz violating parameters $a_{\alpha \beta}$, we also marginalize over their corresponding phases $\phi_{\alpha \beta}$.

 \section*{ Acknowledgements} One of the authors (Rudra Majhi) would like to thank Department of Science \& Technology (DST) Innovation   in Science Pursuit for Inspired Research (INSPIRE) for financial support. The work of RM is supported by SERB, Govt. of India through grant no. EMR/2017/001448. We acknowledge the use of CMSD HPC  facility of Univ. of Hyderabad to carry out computations in this work.

\bibliography{BL}
\end{document}